\documentclass[aps,prb,onecolumn,notitlepage,footinbib,superscriptaddress,longbibliography,showkeys]{revtex4-1}
\usepackage[a4paper,top=2.00cm,bottom=2.00cm,left=2.00cm,right=2.00cm]{geometry}
\setcitestyle{square}
\usepackage[colorlinks=true,linkcolor=blue]{hyperref}
\usepackage{amsmath}
\usepackage{amsfonts}
\usepackage{amssymb}
\usepackage{esint}
\usepackage{mathtools}
\usepackage{bbold}
\usepackage{bm}
\usepackage{graphicx}
\usepackage{color}
\usepackage{multirow}
\usepackage{epstopdf}
\usepackage{ifpdf}
\DeclareGraphicsExtensions{.eps, .pdf, .jpg, .tif}
\usepackage{float}
\def\pder#1#2{\mbox{$\displaystyle\frac{\partial #1}{\partial #2}$}}
\DeclareMathOperator{\Diag}{Diag}
\DeclareMathOperator{\sech}{sech}
\newcommand{\tnsr}[1]{{\overset{\text{\tiny$\boldsymbol\leftrightarrow$}}{#1}}}
\newcommand{\mathtext}[1]{{\mbox{\tiny #1}}}
\def\ns{\negthickspace}
\def\txt#1{\mbox{\tiny {#1}}}
\def\nicefrac#1#2{\genfrac{}{}{}{1}{#1}{#2}}
\def\vsigma{\mbox{$\bm\sigma$}}

\newcommand{\Sig}{{\mbox{\footnotesize $\Sigma$}}}
\newcommand{\grad}{\mbox{$\bm{\nabla}$}}
\newcommand{\ket}[1]{\left|{#1}\right\rangle}
\def\sml{{\bf\textsf{s}}}
\def\tz{{\widehat{\tau}_3}}
\def\ty{{\widehat{\tau}_2}}
\def\tx{{\widehat{\tau}_1}}
\def\tone{\widehat{1}}
\def\tp{{\widehat{\tau}_{+}}}
\def\tm{{\widehat{\tau}_{-}}}
\def\tpm{{\widehat{\tau}_{\pm}}}
\renewcommand{\Re}{\mbox{Re}}
\renewcommand{\Im}{\mbox{Im}}
\def\Tr#1{\mbox{Tr}\left\{#1\right\}}
\def\e{\varepsilon}
\def\vell{\bm{\ell}}
\def\p{\mathbf{\hat p}}
\def\g{\widehat{g}}
\def\vj{\mathbf{j}}
\def\vk{\mathbf{k}}
\def\vg{\mathbf{g}}
\def\vf{\mathbf{f}}
\def\vz{\mathbf{z}}

\def\vd{\mathbf{d}}
\def\vp{\mathbf{p}}

\def\vr{\mathbf{r}}

\def\vv{\mathbf{v}}
\def\vu{\mathbf{u}}
\def\vA{\mathbf{A}}
\def\vB{\mathbf{B}}

\def\vS{\mathbf{S}}

\def\cC{\mathcal{C}}
\def\cY{\mathcal{Y}}
\def\cZ{\mathcal{Z}}
\def\He{$^{3}$He}
\def\Hea{$^{3}$He-A}
\def\upt{UPt$_{3}$}
\def\sro{Sr$_{2}$RuO$_{4}$}

\newcommand{\urs}{URu$_{2}$Si$_{2}$}

\def\imp{\text{imp}}

\def\point#1#2{{\tt #1}_{\mbox{\footnotesize #2}}}
\def\orbital{{\tt SO(3)_{\text{L}}}}
\def\ber{\begin{eqnarray}}
\def\eer{\end{eqnarray}}
\def\be{\begin{equation}}
\def\ee{\end{equation}}
\def\ul{\underline}
\def\sgn{\mbox{\sf sgn}}
\def\bmat{\begin{pmatrix}}
\def\emat{\end{pmatrix}}
\def\la{\left\langle}
\def\ra{\right\rangle}
\setcitestyle{numbers}
\begin{document}
\title{
\vspace*{-2cm}\hspace*{-6cm}
{\footnotesize {\sl Special Issue of Frontiers in Physics: Disorder and Superconductivity: a 21$^{\text{st}}$ century update}}
\\
\vspace*{2cm}
{Anomalous Hall Effects in Chiral Superconductors}
}
\author{Vudtiwat Ngampruetikorn}
\email{vudtiwat.ngampruetikorn@sydney.edu.au}
\affiliation{Center for Applied Physics \& Superconducting Technologies, Department of Physics, Northwestern University, Evanston, IL 60208, USA} 
\affiliation{Fermi National Accelerator Laboratory, Batavia, IL 60510, USA}
\affiliation{Initiative for the Theoretical Sciences, The Graduate Center, CUNY, New York, NY 10016, USA}
\affiliation{School of Physics, University of Sydney, Sydney, NSW 2006, Australia}
\author{{J. A.} {Sauls}}
\email{sauls@lsu.edu}
\affiliation{Center for Applied Physics \& Superconducting Technologies, Department of Physics, Northwestern University, Evanston, IL 60208, USA} 
\affiliation{Fermi National Accelerator Laboratory, Batavia, IL 60510, USA}
\affiliation{Hearne Institute of Theoretical Physics, Department of Physics \& Astronomy, Louisiana State University, Baton Rouge LA 70803, USA}
\begin{abstract}
\medskip
\begin{center}\today\end{center}
\medskip
We report theoretical results for the electronic contribution to thermal and electrical transport for chiral superconductors belonging to even or odd-parity E$_1$ and E$_2$ representations of the tetragonal and hexagonal point groups.
Chiral superconductors exhibit novel properties that depend on the topology of the order parameter and Fermi surface, and---as we highlight---the structure of the impurity potential. An anomalous thermal Hall effect is predicted and shown to be sensitive to the winding number, $\nu$, of the chiral order parameter via Andreev scattering that transfers angular momentum from the chiral condensate to excitations that scatter off the random potential. 
For heat transport in a chiral superconductor with isotropic impurity scattering, i.e., point-like impurities, a transverse heat current is obtained for $\nu=\pm 1$, but vanishes for $|\nu|>1$. This is not a universal result. For finite-size impurities with radii of order or greater than the Fermi wavelength, $R\ge\hbar/p_f$, the thermal Hall conductivity is finite for chiral order with $|\nu|\ge2$, and determined by a specific Fermi-surface average of the differential cross-section for electron-impurity scattering. 
Our results also provide quantitative formulae for analyzing and interpreting thermal transport measurements for superconductors predicted to exhibit broken time-reversal and mirror symmetries.
\end{abstract}
%
\maketitle
\section{Introduction}
The remarkable properties of the spin-triplet, p-wave phases of superfluid \He\ has stimulated research efforts to discover and identify electronic superconductors with novel broken symmetries and non-trivial ground-state topology~\cite{kyc98,mac03,str09,mae12,gan15,avers:20}, driven in part by predictions of novel transport properties. 
\emph{Chiral} superfluids and superconductors are topological phases with gapless Fermionic excitations that reflect the momentum-space topology of the condensate of Cooper pairs.
The A-phase of superfluid \He\ was definitively identified as a chiral p-wave superfluid by the observation of anomalous Hall transport of electrons moving through a quasiparticle fluid of chiral Fermions \cite{ike13,she16}.
A chiral d-wave state was proposed for doped graphene~\cite{nan12,bla15}, while a chiral p-wave state is proposed for MoS~\cite{yua16}.
There is evidence from $\mu$SR of broken time-reversal symmetry onsetting at the superconducting transition for the pnictide SrPtAs~\cite{bis13}, and a chiral d-wave state has been proposed as the ground state~\cite{fis14}. Recent $\mu$SR experiments also provide evidence for chiral d-wave superconductivity in the pnictide LaPt\textsubscript{3}P~\cite{biswas:21}.
The perovskite, \sro, has been studied extensively and was proposed as a promising candidate for chiral p-wave superconductivity ($\point{E}{u}$ pairing with $\vec{\Delta}=\hat\vd\,(p_x+ip_y)$), in part based on similarities of its normal-state Fermi-liquid properties with those of liquid \He~\cite{ric95,kal12}.
Evidence of broken time-reversal symmetry from both $\mu$SR and Kerr rotation measurements support an identification of \sro\ as a chiral superconductor~\cite{luk98,xia06}. However, experiments designed to detect the theoretically predicted chiral edge currents~\cite{mat99}, or to test for the two-dimensionality of the $\point{E}{u}$ representation that is a necessary requirement to support a chiral ground state, are so far inconclusive, or report null results~\cite{kir07,hic10,cur14,hic14}.
Recent transport measurements also appear to conflict with the chiral p-wave identification based on the $\point{E}{u}$ representation; i.e. thermal conductivity measurements at low temperatures and as a function of magnetic field, which probe the low-energy quasiparticle excitation spectrum, are consistent with the nodal line structure of a d-wave order parameter, and inconsistent with the gap structure expected based on the $\point{E}{u}$ representation~\cite{has17,gra00b}. The possibility that \sro\ is an even parity chiral superconductor has so far not been ruled out (see, e.g.\ Refs.~\cite{mackenzie:17,kashiwaya:19,pustogow:19,sharma:20,chronister:21}).

The first superconductor reported to show experimental evidence of broken time-reversal symmetry was the heavy fermion superconductor, \upt, based on $\mu$SR linewidth measurements~\cite{luk93}. This experiment followed theoretical predictions of broken time-reversal symmetry in the B-phase of \upt, i.e.\ the lower temperature superconducting phase~\cite{hes89}. Another notable signature of broken time-reversal symmetry is the onset of Kerr rotation as \upt\ enters its low-temperature B-phase~\cite{sch14}. More recently, a neutron scattering experiment, using vortices as a probe for the superconducting state in the bulk, offers yet another piece of evidence for broken time-reversal symmetry in \upt~\cite{avers:20}. 
These results support the identification of a chiral superconducting phase of \upt, and they also support the basic theoretical model of a multi-component order parameter belonging to a two-dimensional representation of the hexagonal point group, $\point{D}{6h}$, in which a weak symmetry breaking field lifts the degeneracy of the two-component order stabilizing two distinct superconducting phases in zero magnetic field~\cite{hes89,mac89}.
In this theory the predicted A phase of \upt\ is time-reversal symmetric with pronounced anisotropic pairing correlations in the hexagonal plane~\cite{hux00,str10}, is preferentially selected by the symmetry breaking field, and nucleates at $T_{c_1} = 560\,\mbox{mK}$ as the first superconducting phase.
The B-phase develops as the sub-dominant partner of the two-dimensional representation nucleates at $T_{c_2}\approx 470\,\mbox{mK}$, such that the low-temperature superconducting phase spontaneously breaks both time-reversal and mirror-reflection symmetries, the latter defined by a plane containing the chiral axis which is aligned parallel (or anti-parallel) to the c-axis of \upt.

There are four two-dimensional representations of $\point{D}{6h}$: two even-parity representations, $\point{E}{1g}$ and $\point{E}{2g}$, and two odd-parity representations, $\point{E}{1u}$ and $\point{E}{2u}$, all of which allow for chiral ground states~\cite{sau94,joy02}.
The chiral ground states belonging to the $\point{E}{1}$ and $\point{E}{2}$ representations are defined by zeroes of the Cooper pair amplitude at points $\vp = \pm p_f\,\hat\vz$ on the Fermi surface that are protected by the topology of the orbital order parameter in momentum space, i.e. $\Delta(\vp) = |\Delta(\vp)|\,e^{i\nu\,\phi_{\vp}}$, where $\phi_{\vp}$ is the azimuthal angle defining a point $\vp$ on the Fermi surface, and $\nu = \pm 1$ ($\nu = \pm 2$) for the $\point{E}{1}$ ($\point{E}{2}$) representations~\footnote{More complex chiral order parameters with large winding numbers are allowed by the point group symmetry, c.f. Ref.~\cite{sca15}; $\nu=\pm 1,\pm 2$ are the loweset order harmonics consistent with the $\point{E}{1}$ and $\point{E}{2}$ representations, respectively.}
The bulk of the experimental evidence---thermodynamic, H-T phase diagram~\cite{cho91,cho93,sau94}, thermal transport~\cite{gra96b,gra99}, ultra-sound~\cite{gra00a}, Josephson tunneling~\cite{str09}, SANS~\cite{gan15} and optical spectroscopy measurements~\cite{sch14}---supports the identification of \upt\ as an odd-parity superconductor with an order parameter belonging to the $\point{E}{2u}$ representation, \emph{and} a chiral B-phase order parameter of the form, $\vec{\Delta}_{\pm}(\vp)= \Delta_{\mbox{\tiny B}}(T)\,\hat\vd\, \hat{p}_z\left(\hat{p}_x\pm i\hat{p}_y\right)^2\,\sim\hat\vd\,e^{\pm i\,2\phi_{\vp}}$.  The vector $\hat\vd$ is the quantization axis along which the spin-triplet Cooper pairs have zero spin projection, i.e. an equal-spin pairing (ESP) state~\cite{cho91}.
A key feature of the $\point{E}{2u}$ chiral order parameter is the winding number $\nu=\pm 2$. The Josephson interference experiment described in Ref.~\cite{str09} can discriminate between $|\nu|=1$ and $|\nu|=2$ chiral ground states. Indeed the report of a $\pi$ phase shift in the Fraunhofer pattern for the corner-SQUID geometry, combined with the observations of broken time-reversal symmetry~\cite{luk93,sch14}, provides strong evidence in favor a $|\nu| = 2$ ($\point{E}{2u}$) chiral B-phase of \upt. However, conclusive evidence for \emph{bulk} chiral superconductivity remains elusive. A zero-field Hall transport measurement is an ideal experiment to confirm broken time-reversal and mirror symmetries in the bulk of a chiral superconductor candidate.

\vspace*{-7mm}
\section{Anomalous Hall Transport}

The winding number of the order parameter for a chiral superconductor reflects the topology of the superconducting ground state. For a fully gapped chiral superconductor $\nu$ is related to the Chern number defined in terms of the Bogoliubov-Nambu Hamiltonian in 2D momentum space, or for chiral superconductors defined on a 3D Fermi surface the effective two-dimensional spectrum at fixed $p_z\ne 0$, $\cC(p_z) = \nicefrac{1}{2\pi}\int d^2p\,\Omega_z({\vp})$, where $\Omega_z(\vp)$ is the Berry curvature~\cite{gos15}. The result for the Chern number is $\cC(p_z) = \nu$, which provides topological protection for a spectrum of chiral Fermions.

For 2D chiral phases there is a spectrum of massless chiral Fermions confined on the boundary (edge states) with the zero-energy state enforced by the bulk topology. However, for a chiral order parameter defined on a closed 3D Fermi surface there is also a bulk spectrum of gapless Weyl-Majorana Fermions with momenta near the nodal points $p_z = \pm p_f$, in addition to a \emph{spectrum} of massless chiral Fermions confined on surfaces normal to the [1,0,0] and [0,1,0] planes~\cite{tsu12}.

\subsection{Anomalous Edge Transport}

For a fully gapped chiral p-wave ground state in two dimensions the spectrum of chiral edge Fermions is predicted to give rise to quantized heat and mass transport in chiral superfluids and superconductors~\cite{rea00,sto04,sau11,tsu12,sum13}.
In particular, an anomalous thermal Hall conductance is predicted to be quantized, $K_{xy}/k_{\mathtext{B}} T = \nicefrac{\pi}{12}\,k_{\mathtext{B}}/\hbar$ based on the low-energy effective field theory of the chiral edge states~\cite{rea00,qin11}. This result is also obtained from the topology of the bulk order parameter combined with linear response theory based on the Bogoliubov Hamiltonian for 2D $p_x+ip_y$ topological superconductors~\cite{sum13}.

For a chiral superconductor defined on a 3D Fermi surface an anamolous thermal Hall current is predicted, but is not quantized in units of a fundamental quantum of conductance. Based on the linear response theory of Qin et al.~\cite{qin11} Goswami and Nevidomsky obtained a result for the anomalous thermal Hall conductivity of the B-phase of \upt\ for $T\ll T_{c_2}$~\cite{gos15},
\be
\kappa_{xy} / k_{\mathtext{B}} T = \nu\,\frac{\pi}{6}\, 
\frac{k_{\mathtext{B}}}{\hbar}\,\left(\frac{\Delta p}{2\pi\hbar}\right) 
\,.
\ee
The anomalous thermal Hall conductivity reflects the number of branches of chiral Fermions confined on the [1,0,0] or [0,1,0] surface, i.e. $|\nu| = 2$ for the $\point{E}{2u}$ chiral ground state. 
The non-universality of the thermal Hall conductivity is reflected by the term $\Delta p$, which is the ``distance'' between the two topologically protected $\nu=2$ Weyl points at $\hat{p}_z=\pm 1$ on a projected surface containing the chiral axis; e.g.\ $\Delta p = 2p_f$ for a spherical Fermi surface~\cite{gos15}.

Thus, heat transport experiments could decisively identify the broken symmetries and topology of superconductors predicted to exhibit chiral order. The thermal conductivity depends on both the topology of the order parameter and the Fermi surface. The anomalous thermal Hall effect, in which a temperature gradient generates heat currents perpendicular to it, results from broken time-reversal and mirror symmetries---a direct signature of chiral pairing.~\footnote{Note that time-reversal and mirror symmetries need not be broken simultaneously. For example, a three-band superconductor may break time-reversal symmetry when the order parameter defined on each band has a different phase~\cite{sta10}. Mirror symmetry is however preserved and Hall effects are therefore not expected in this system.}
A zero-field thermal Hall experiment can also be used as a signature of chiral edge states. However, zero-field thermal Hall transport has remained elusive thus far.

\subsection{Impurity-Induced Anomalous Transport}

Here we consider zero-field Hall transport resulting from electron-impurity interactions in the bulk of the superconductor, which we show are easily several orders of magnitude larger than the edge contribution~\cite{wave:20}. There are earlier theoretical predictions for impurity-induced anomalous thermal Hall effects in chiral superconductors based on point-like impurities by several authors~\cite{Arfi:1988,Li:2015,Yip:2016}.
As we show, the point-like impurity model, which includes only s-wave quasiparticle-impurity scattering, predicts zero Hall response except for Chern number $\nu=\pm 1$~\cite{wave:20}, i.e. only for chiral p-wave superconductors~\cite{Arfi:1988}.

In the following we present a self-consistent theory incorporating the effects of finite-size impurities and show that such effects are essential for a quantitative description of Hall transport in chiral superconductors. 
Experimental observation of an impurity-induced anomalous thermal Hall effect would provide a definitive signature of chiral superconductivity. The bulk effect can easily dominate the edge state contribution to the anomalous Hall current, except in ultra-pure fully gapped chiral superconductors.

\section{Transport Theory}

We start from the Keldysh extension~\cite{kel65} of the transport-like equations originally developed by Eilenberger, Larkin and Ovchinnikov for equilibrium states of superconductors~\cite{eil68,lar69}, and extended by Larkin and Ovchinnikov to describe superconductors out of of equilibrium~\cite{lar75}. This formalism is referred to as ``quasiclassical theory''. For reviews see Refs.~\cite{ser83,lar86,rai94b}.
The quasiclassical theory is formulated in terms of $4\times 4$ matrix propagators for Fermionic quasiparticles and Cooper pairs that describe the space-time evolution of the their non-equilibrium distribution functions, as well as the dynamical response of the low-energy spectral functions and the superconducting order parameter.
Here we are interested in the response to static, or low-frequency, thermal gradients and external forces that couple to energy, mass and charge currents. We follow as much as possible the notation and conventions of theory developed for thermal transport in unconventional superconductors by \citet{gra96a}.

\subsection{Keldysh-Eilenberger Equations}

The quasiclassical transport equations are matrix equations in particle-hole (Nambu) space which describe the dynamics of quasiparticle excitations and Cooper pairs. Physical properties, such as the spectral density, currents or response functions are expressed in terms of components of the Keldysh matrix propagator,
\be
\check{G}(\vp,\varepsilon;\vr,t)=
\begin{pmatrix} \widehat{g}^R & \widehat{g}^K \cr 0 & \widehat{g}^A \end{pmatrix}
\,,
\ee
where $\widehat{g}^{R,A,K}(\vp,\varepsilon;\vr,t)$ are the $4\times 4$ retarded (R), 
advanced (A) and Keldysh (K) matrix propagators.

The nonequilibrium dynamics is described by a transport equation for the Keldysh propagator,
\begin{eqnarray}\label{QC_K}
\widehat{H}^{R}\circ\widehat{g}^{K}(\vp,\vr;\varepsilon,t)
-\widehat{g}^{K}\circ\widehat{H}^{A}(\vp,\vr;\varepsilon,t)
+\widehat{g}^{R}\circ\widehat{\Sig}^{K}(\vp,\vr;\varepsilon,t)
-\widehat{\Sig}^{K}\circ\widehat{g}^{A}(\vp,\vr;\varepsilon,t)
+i\vv_{\vp}\cdot\grad\widehat{g}^{K}(\vp,\vr;\varepsilon,t) = 0
\,,
\end{eqnarray}
as well as transport equations for the retarded and advanced propagators,
\be\label{QC_RA}
\left[\widehat{H}^{R,A}\,,\,\widehat{g}^{R,A}\right]_{\circ}
 +i\vv_{\vp}\cdot\grad\widehat{g}^{R,A}(\vp,\vr;\varepsilon,t) = 0
\,,
\ee
where
\be\label{QCoperator}
\widehat{H}^{R,A}(\vp,\vr;\varepsilon,t)
= 
\varepsilon\tz-\widehat{v}(\vp,\vr;t)-\widehat{\Sig}^{R,A}(\vp,\vr;\varepsilon,t)
\ee
is defined in terms of the excitation energy, $\varepsilon$, the coupling to external fields, 
$\widehat{v}$, and the self-energies, $\widehat{\Sig}^{R,A}$. Pairing correlations, as well as 
effects of scattering by impurities, phonons and quasiparticles are described by the 
self-energies, $\widehat{\Sig}^{R,A,K}$. 
The convolution product ($\circ$-product) appearing in Eqs. (\ref{QC_K}-\ref{QC_RA}), in the 
mixed energy-time representation, is defined by,
\be\label{convolution}
\widehat{A}\circ\widehat{B}(\varepsilon;t)= e^{\frac{i}{2}
\left[{\partial^{A}_{\varepsilon}\partial^{B}_{t}-
       \partial^{A}_{t}\partial^{B}_{\varepsilon}}\right]}\,
\widehat{A}(\varepsilon;t)\widehat{B}(\varepsilon;t)
\,.
\ee
Note that $\varepsilon$ is the excitation energy and $t$ is the external time variable.  The operator expansion for the convolution product is particularly useful if the external timescale, $t\sim\omega^{-1}$ is slow compared to the typical internal dynamical timescales, $\hbar/\Delta$ and $\tau$, i.e. $\omega\ll|\varepsilon|\sim\Delta$ and $\omega\ll 1/\tau$. In this limit we can expand Eq.~(\ref{convolution}),
\be\label{convolution-low-frequency}
\widehat{A}\circ\widehat{B}(\varepsilon;t)\approx
\widehat{A}(\varepsilon;t)\widehat{B}(\varepsilon;t)
+\frac{i}{2}
\left[\pder{\widehat{A}}{\varepsilon}\pder{\widehat{B}}{t} 
-     \pder{\widehat{A}}{t}\pder{\widehat{B}}{\varepsilon}\right]
\,.
\ee
The quasiclassical transport equations are supplemented by the normalization conditions~\cite{eil68,lar69},
\begin{align}\label{normRA}
\widehat{g}^{R,A}\circ\widehat{g}^{R,A} &= -\pi^2\,\tone
\\
\widehat{g}^{R}\circ\widehat{g}^{K}-\widehat{g}^{K}\circ\widehat{g}^{A}&=0.
\label{normK}
\end{align}

\subsection{Quasiclassical Propagators}

The quasiclassical propagators are $4\times 4$-matrices whose structure describes the internal quantum-mechanical degrees of freedom of quasiparticles and quasiholes.  In addition to spin, the particle-hole degree of freedom is of fundamental importance to our understanding of superconductivity. In the normal state of a metal or Fermi liquid there is no quantum-mechanical coherence between particle and  hole excitations. By contrast, the distinguishing feature of the superconducting state is the existence of quantum mechanical coherence between normal-state particle and hole excitations. Particle-hole coherence is the origin of persistent currents, Josephson effects, Andreev scattering, flux quantization, and all other nonclassical superconducting effects. The quasiclassical propagators are directly related to density matrices which describe the quantum-statistical state of the internal degrees of freedom. Nonvanishing off-diagonal elements in the particle-hole density matrix are indicative of superconductivity, indeed the onset of non-vanishing off-diagonal elements is the signature of the superconducting transition. 

The Nambu matrix structure of the propagators and self energies is 
\be\label{gmatrx}
\widehat{g}^{R,A,K}
=
\begin{pmatrix}
g^{R,A,K} + \vg^{R,A,K}\cdot\vsigma
& 
\left(f^{R,A,K} + \vf^{R,A,K}\cdot\vsigma\right)i\sigma_y
\cr
i\sigma_y\left(\bar{f}^{R,A,K}+ \bar{\vf}^{R,A,K}\cdot\vsigma\right)
&
\bar{g}^{R,A,K}- \bar{\vg}^{R,A,K}\cdot\sigma_y\vsigma\sigma_y
\end{pmatrix}
\,.
\ee
The 16 matrix elements of $\widehat{g}^{R,A,K}$ are expressed in terms of 4 spin-scalars 
($g^{R,A,K}$, $\bar{g}^{R,A,K}$, $f^{R,A,K}$, $\bar{f}^{R,A,K}$) and 4 spin-vectors 
($\vg^{R,A,K}$, $\bar{\vg}^{R,A,K}$, $\vf^{R,A,K}$, $\bar{\vf}^{R,A,K}$). All matrix 
elements are functions of $\vp$, $\varepsilon$, $\vr$ and $t$. The spin scalars $g^{R,A,K}$, 
$\bar{g}^{R,A,K}$ determine spin-independent properties such as the charge, mass and heat 
current densities, 
$\vj_e(\vr,t)$, $\vj_m(\vr,t)$ and $\vj_q(\vr,t)$, 
as well as the local density of states
\be\label{local-DOS}
\hspace*{-2mm}
N(\varepsilon;\vr,t)=N_f\int\ns d\vp
\left[-\frac{1}{\pi}\Im\,\nicefrac{1}{2}\Tr{\tz\,\widehat{g}^R(\vp,\varepsilon;\vr,t)}\right]
\,,
\ee
where $N_f$ is the normal-state density of states at the Fermi energy. The integration 
is over the Fermi surface {\sl weighted} by the angle-resolved normal density of states 
at the Fermi surface, $n(\vp)$, normalized to 
\be\label{FS_average}
\hspace*{-4mm}
\int d\vp \left(...\right)\equiv\int\,dS_{\vp}n(\vp)\left(\ldots\right)
\;\;
\text{with}
\;\;
\int\,dS_{\vp}n(\vp) = 1
\,.
\ee

The current densities are determined from Fermi-surface averages over the elementary currents, $[e\vv_{\vp}]$, mass, $[m\vv_{\vp}]$, and energy, $[\varepsilon\vv_{\vp}]$, weighted by the scalar components of the diagonal Keldysh propagator. 
In particular, the charge and heat current densities are given by 
\begin{align}
\vj^{(e)}(\vr,t) &= N_f\ns\int\ns d\vp\ns\int\frac{d\varepsilon}{4\pi i}
\left[e\vv_{\vp}\right]
\Tr{\tz\,\widehat{g}^K(\vp,\varepsilon;\vr,t)}\,,
\label{charge-current}
\\
\vj^{(q)}(\vr,t) &= N_f\ns\int\ns d\vp\ns\int\frac{d\varepsilon}{4\pi i}
\left[\varepsilon\vv_{\vp}\right]
\Tr{\widehat{g}^K(\vp,\varepsilon;\vr,t)}\,.
\label{heat-current}    
\end{align}

The off-diagonal components, $f^{R,A,K}$ and $\vf^{R,A,K}$, are the anomalous propagators 
that characterize the pairing correlations of the superconducting state. 
Spin-singlet pairing correlations are encoded in $f^{K}$, while $\vf^{K}$ is the measure of
spin-triplet pairing correlations. Pair correlations develop spontaneously at temperatures below the 
superconducting transition temperature $T_c$. The anomalous propagators are not directly 
measurable, but the correlations they describe are observable via their coupling 
to the ``diagonal'' propagators, $g^{R,A,K}$ and $\vg^{R,A,K}$, through the transport equations.

\subsection{Coupling to External and Internal Forces}

The couplings of low-energy excitations to electromagnetic fields are defined in terms of the  
scalar and vector potentials,
\be
\widehat{v}_{\mathrm{EM}} = e\,\varphi(\vr,t)\tz + \frac{e}{c}\vv_{\vp}\cdot\vA(\vr,t)\tz
\,.
\ee
Note that $e\tz$ encodes the charge coupling of both particle and hole excitations to the 
electromagnetic field. The magnetic field also couples to the quasiparticles and pairs via 
the Zeeman energy, $\widehat{v}_{\txt{Z}} = \gamma\,\widehat{\vS}\cdot\vB(\vr,t)$,
where $\vB=\grad\times\vA$, $\gamma$ is the gyromagnetic ratio of the normal-state quasiparticles, 
and 
$\widehat{\vS}=\nicefrac{1}{2}(\tone+\tz)\vsigma-\nicefrac{1}{2}(\tone-\tz)\sigma_y\vsigma\sigma_y$
is the Nambu representation of the Fermion spin operator.

\subsubsection*{Mean-Field Self-Energies}

Superconductors driven out of equilibrium are also subject to internal forces on quasiparticles and Cooper pairs, originating from electron-electron, electron-phonon and electron-impurity interactions. These interactions enter the quasiclassical theory as self-energy terms, $\widehat{\Sig}^{R,A,K}$, in the transport Eqs.~\eqref{QC_K}, \eqref{QC_RA} and \eqref{QCoperator}.
We include self-energies that contribute to leading order in expansion parameters, $\sml=\{1/k_f\xi,\,k_{B}T_c/E_f,1/k_f\ell,\hbar/\tau E_f$, $\Delta/E_f\,\ldots\}\ll 1$, that define the low-energy, long-wavelength region of validity of Landau Fermi-liquid theory, and its extension to include BCS condensation~\cite{ser83,rai86,rai94b}.

The leading order contributions to the self-energy from quasiparticle-quasiparticle interactions correspond the mean-field self-energies, $\widehat\Sigma^{R,A,K}_{\text{mf}}$, in the particle-hole (Landau) and particle-particle (Cooper) channels, and are represented by Eqs.~\eqref{eq-Sigma_Landau} and \eqref{eq-Sigma_Cooper}, respectively~\footnote{The sign in Eq.~\eqref{eq-Sigma_Cooper} is such that $\lambda^{s,\,t} > 0$ corresponds to an attractive pairing interaction.},
\begin{align}
\label{eq-Sigma_Landau}
\hat\Sigma(\vp)
&=
\int\ns d\vp'\,\fint\ns\frac{d\varepsilon'}{4\pi i}
\left[
A^{s}(\vp,\vp')\,g^{K}(\vp',\varepsilon')\,
\hat{1}
+
A^{a}(\vp,\vp')\,\vg^{K}(\vp',\varepsilon')\cdot
\vsigma
\right]
\,,
\\
\hat\Delta(\vp) 
&=
\int\ns d\vp'\fint\ns\frac{d\varepsilon'}{4\pi i} 
\left[
\lambda^{s}(\vp,\vp')\,f^{K}(\vp',\varepsilon')\,i\sigma_y 
+
\lambda^{t}(\vp,\vp')\,\vf^{K}(\vp',\varepsilon')\cdot i\vsigma\sigma_y
\right]
\,.
\label{eq-Sigma_Cooper}
\end{align}
Note that $\hat\Sigma$ and $\hat\Delta$ represent the upper row of the Nambu matrix, $\widehat\Sigma_{\text{mf}}$. Since the mean-field self-energies are independent of $\varepsilon$, $\widehat\Sigma_{\text{mf}}^R = \widehat\Sigma_{\text{mf}}^A=\widehat\Sigma_{\text{mf}}$, and $\widehat\Sigma_{\text{mf}}^K=0$. 
The interaction vertex, $A(\vp,\vp')$, in Eq.~\eqref{eq-Sigma_Landau} represents the quasiparticle interactions in the particle-hole channel.
In the non-relativistic limit these interactions are spin-rotation invariant, in which case there are two real amplitudes: the spin-independent quasiparticle-quasiparticle interaction, $A^{s}(\vp,\vp')$, the \emph{exchange} term, $A^{a}(\vp,\vp')$, describing the spin-dependent quasiparticle-quasiparticle interaction. These interactions are defined by the renormalized four-point vertex in the forward-scattering limit for quasiparticles with momenta and energies confined to the Fermi surface, i.e. $|\vp| = |\vp'|=p_f$ and $\varepsilon=\varepsilon'=0$, which is a good approximation in the Fermi-liquid regime far from a quantum critical point. 
Thus, the propagator is integrated over the low-energy bandwidth defined by $\fint(\ldots)\equiv\int_{-\varepsilon_c}^{+\varepsilon_c}(\ldots)$, and the corresponding self-energies depend on the direction of the quasiparticle momentum on the Fermi surface, but are independent of $\varepsilon$.

In the Cooper channel the mean-field self energy from quasiparticle interactions is given by Eq.~\eqref{eq-Sigma_Cooper}. The interaction vertex separates in terms of an even-parity, spin-singlet interaction, $\lambda^{s}(\vp,\vp')$, and an odd-parity, spin-triplet interaction, $\lambda^{t}(\vp,\vp')$, the latter resulting from exchange symmetry in the non-relativistic limit.\footnote{This separation does not apply to superconducting materials without an inversion center, i.e. non-centrosymmetric superconductors.}
In a rotationally invariant Fermi liquid like liquid \He, the interactions in the Cooper channel further separates according to the irreducible representations of the rotation group in three dimensions, $\orbital$,
\be\label{eq-Cooper-channel_irreps}  
\lambda^{s\,(t)}(\vp,\vp')=\sum_{l}^{\text{even (odd)}}\lambda_l
		    \sum_{m=-l}^{+l}\ns\cY^*_{l,m}(\hat\vp)\cY_{l,m}(\hat\vp')
\,,
\ee
which are labeled by the orbital angular momentum quantum number $l\in\{0,1,2,\ldots\}$, with the basis functions given by the spherical harmonics $\{\cY_{lm}(\hat\vp)\}$, normalized to $\int d\vp\,\cY_{lm}(\hat\vp)\cY_{l'm'}(\hat\vp') = \delta_{ll'}\delta_{mm'}$.
The Cooper instability occurs in the pairing channel defined by the most attractive interaction, $\lambda_l$, which for \He\ is the odd-parity, spin-triplet ($S=1$), $l=1$ (p-wave) channel.
For strongly correlated materials Cooper pairing is mediated by quasiparticle-quasiparticle interactions. This is necessarily the case in a single-component Fermi system like liquid \He, and is prevalent in strongly correlated electronic compounds such as the heavy-fermion superconductors, \upt\ and \urs, and unconventional superconductors like \sro, all of which exhibit experimental signatures of broken time-reversal symmetry by the superconducting state.
For these superconductors the pairing channel belongs to an irreducible representation of the crystal point group. Equation \eqref{eq-Cooper-channel_irreps} holds with $l$ summed over the irreducible representations of the point group, the second sum $m$ is over the set of orthogonal basis functions, $\{\cY_{lm}(\vp) | m \in \mbox{irrep}_l\}$, that span the irrep labeled by $l$.
For materials with hexagonal point symmetry, e.g.\ \upt, we consider the four two-dimensional ``E-reps'': even parity $\point{E}{1g}$ and $\point{E}{2g}$ representations and odd-parity $\point{E}{1u}$ and $\point{E}{2u}$. All four E-reps allow for a chiral ground state with minimum Chern numbers of $\nu = \pm 1$ ($\point{E}{1g(u)}$) or $\nu=\pm 2$ ($\point{E}{2g(u)}$).

\subsubsection*{Impurity Self-Energy}

The effects of impurity disorder originate from the quasiparticle-impurity interaction, $\check{u}(\vp,\vp')$, which corresponds to the transition matrix element for elastic scattering of a quasiparticle with momentum $\vp$ to the point $\vp'$ on the Fermi surface.
Multiple scattering of quasiparticles and quasiholes by an impurity is described by the Bethe-Salpeter equation,
\be
\check{t}(\vp',\vp;\epsilon) 
=
\check{u}(\vp',\vp) 
+ N_f\ns\int\ns d\vp''\,
\check{u}(\vp',\vp'')\,
\check{g}(\vp'';\epsilon)
\check{t}(\vp'',\vp;\epsilon) 
\,,
\label{eq-t_matrix}
\ee
where $\check{t}(\vp',\vp;\varepsilon)$ is the t-matrix for quasiparticle-impurity scattering, and $\check{g}(\vp;\varepsilon)$ is the quasiclassical Keldysh matrix propagator for particles, holes and Cooper pairs.
The leading-order contribution to the configurational-averaged self energy is then determined by scattering of quasiparticles off an uncorrelated, random distribution of statistically equivalent impurities with average density, $n_{\text{imp}}$, 
\ber
\check{\Sig}_{\text{imp}}(\vp;\epsilon)
=
n_{\text{imp}}\,\check{t}(\vp,\vp;\epsilon)
=
\begin{pmatrix} 
\widehat{\Sig}^{R}_{\text{imp}}
&
\widehat{\Sig}^{K}_{\text{imp}}
\\
0
&
\widehat{\Sig}^{A}_{\text{imp}}
\end{pmatrix},
\label{fmf-Sigma_impurity}
\eer
where $\check{t}(\vp,\vp;\varepsilon)$ is the t-matrix evaluated self-consistently in the forward-scattering limit. Thus, the Nambu-matrix components of the impurity Keldysh self energy, $\widehat{\Sig}_{\text{imp}}^{R,A,K}(\vp,\vr;\varepsilon,t)=n_{\text{imp}}\,\widehat{t}^{R,A,K}(\vp,\vp,\vr;\varepsilon,t)$, are determined by the corresponding components of the t-matrix,
\begin{align}
    \label{eq-tmatrix-RA} 
\widehat{t}^{R,A}(\vp',\vp,\vr;\varepsilon,t) 
&= u(\vp',\vp) + N_f\int d\vp''\,u(\vp',\vp'')\,
  \widehat{g}^{R,A}(\vp'',\vr;\varepsilon,t)\circ
  \widehat{t}^{R,A}(\vp'',\vp,\vr;\varepsilon,t)
\\
\label{eq-tmatrix-K} 
\widehat{t}^K(\vp',\vp,\vr;\varepsilon,t)
&= N_f\int d\vp''\,\widehat{t}^R(\vp',\vp'',\vr;\varepsilon,t)\circ
              \widehat{g}^K(\vp'',\vr;\varepsilon,t)\circ
	      \widehat{t}^A(\vp'',\vp,\vr;\varepsilon,t)
\,.
\end{align}
Before proceeding to non-equilibrium quasiparticle transport we need to discuss the equilibrium state, including the effects of impurity scattering, on the equilibrium states of chiral superconductors and superfluids.

\section{Equilibrium}

For homogeneous systems in equilibrium the transport equations for the retarded and advanced propagators reduce to
\be\label{eq-QC-RA_equilibrium}
\left[\varepsilon\tz-\widehat{\Delta}(\vp)-\widehat\Sig_{\text{imp}}^{R,A}(\vp;\varepsilon)
\,,\,
\widehat{g}^{R,A}(\vp;\varepsilon)\right] = 0
\,,
\ee
where $\widehat\Delta(\vp)$ is the mean-field order parameter and $\widehat\Sig_{\text{imp}}^{R,A}(\vp;\varepsilon)$ are the equilibrium self-energies resulting from quasiparticle-impurity scattering.
We consider the low-temperature limit in which the thermal populations of quasiparticles and phonons are sufficiently small that we can neglect quasiparticle-quasiparticle scattering and quasiparticle-phonon scattering contributions to the self energy. Thus, we retain only the mean-field pairing self energy and impurity self energy resulting from elastic quasiparticle-impurity scattering. The propagator is also constrained by the normalization condition, which for equilibrium reduces to matrix multiplication,
\be\label{eq-normalization-equilibrium}
\left[\widehat{g}^{R,A}(\vp,\varepsilon)\right]^2 = -\pi^2\,\tone
\,.
\ee

A chiral superconducting ground state is defined by spontaneous breaking of time-reversal and mirror symmetries by the orbital state of the Cooper pairs. We restrict our analysis to \emph{unitary} superconductors in which the $4\times 4$ Nambu matrix order parameter obeys the condition,
\be
\widehat{\Delta}(\vp)^2 = -|\Delta(\vp)|^2\,\tone
\,.
\ee
Unitary states preserve time-reversal symmetry with respect to the spin-correlations of the pairing state. In the clean limit $|\Delta(\vp)|$ is the energy gap for quasiparticles with momentum $\vp$ near the Fermi surface, i.e. the Bogoliubov quasiparticle excitation energy is doubly degenerate with respect to spin and given by $E_{\vp} = \sqrt{\xi_{\vp}^2 + |\Delta(\vp)|^2}$, with $\xi_{\vp}  = v_f(|\vp|-p_f)$ and $\Delta(\vp)$ defined for $\vp$ on the Fermi surface.
The unitarity condition is necessarily satisfied by spin-singlet pairing states, and is also the case for all known spin-triplet superconductors in which the parent state in zero external field is non-magnetic~\footnote{The $A_{1}$ phase of superfluid \He, which is stabilized by an externally applied magnetic field, is a non-unitary spin-triplet state~\cite{amb73}. The Uranium-based ferromagnetic superconductors are also believed to be non-unitary, spin-polarized, triplet superconductors.}. 
An important class of unitary triplet states are the equal-spin-pairing (ESP) states defined by the $2\times 2$ spin-matrix order parameter, $\hat\Delta(\vp) = \Delta(\vp)\,\hat\vd\cdot(i\vsigma\sigma_y)$, in which $\hat\vd$ is the direction in spin space along which the Cooper pairs have zero spin projection. Equivalently, this state corresponds to equal amplitudes for the spin projections $S_u=+1$ and $S_u=-1$ with $\hat\vu\perp\hat\vd$. For the chiral A-phase of \He, the direction $\hat\vd$ can be controlled by a small magnetic field, $\vB$, through the nuclear Zeeman energy that orients $\vd\perp\vB$. For chiral superconductors spin-orbit coupling and the crystalline field typically lock $\vd$ along a high-symmetry direction of the crystal. 

We consider four classes of chiral ground states corresponding to the even-parity, spin-singlet, $\point{E}{1g}$ and $\point{E}{2g}$, and odd-parity, spin-triplet, $\point{E}{1u}$ and $\point{E}{2u}$ representations of the hexagonal point group, $\point{D}{6h}$. These representations all allow for chiral ground states with principle winding numbers, $\nu=\pm 1$ ($\nu=\pm 2$) for the $\point{E}{1}$ ($\point{E}{2}$) representations~\footnote{For a cylindrical Fermi surface with \emph{continuous} 2D rotational symmetry, $\point{D}{\mbox{$\infty$}h}$, chiral ground states with any integer winding number $\nu\in\cZ$ are possible. For the discrete point group $\point{D}{6h}$ higher winding numbers with $\nu=\pm 1+\mbox{mod}(6)$ ($\point{E}{1}$) or $\nu = \pm 2 + \mbox{mod}(6)$ ($\point{E}{2}$) are possible for pairing basis functions exhibiting strong hexagonal anisotropy, but in general the chiral basis functions with higher winding numbers will mix with $\nu=\pm 1$ ($\nu=\pm 2$).}. 
Table \ref{table-Ereps} provides representative basis functions for these two-dimensional representations.

\begin{table}
\centering
\begin{tabular}{ccr@{$\,\propto\,$}lcc}
\hline
\begin{tabular}{c}Point\\[-1ex]Group\end{tabular}	&
\begin{tabular}{c}Irrep\\[-1ex]$\Gamma$\end{tabular}	&
\multicolumn{2}{c}{
\begin{tabular}{c}
Basis functions\\[-1ex]$\eta_{\Gamma,\nu}(\p)$
\end{tabular}
}	&
\begin{tabular}{c}Winding\\[-1ex]number $\nu$\end{tabular}
&
Parity
\\
\hline
\multirow{2}{*}{$D_{4h}$}	
& $E_g$	&$\hat p_z(\hat p_x\pm i\hat p_y)$ & $\cY^{\pm 1}_2(\p)$ & $\pm 1$ & $+$ \\
& $E_u$	&$\hat p_x\pm i\hat p_y$         &   $\cY^{\pm 1}_1(\p)$ & $\pm 1$ & $-$ \\
\hline
\multirow{4}{*}{$D_{6h}$}
& $E_{1g}$ &$\hat p_z(\hat p_x\pm i\hat p_y)$ &	$\cY^{\pm 1}_2(\p)$ & $\pm 1$ & $+$ \\
& $E_{2g}$ &$(\hat p_x\pm i\hat p_y)^2$       &	$\cY^{\pm 2}_2(\p)$ & $\pm 2$ & $+$ \\
& $E_{1u}$ &$\hat p_x\pm i\hat p_y$           &	$\cY^{\pm 1}_1(\p)$ & $\pm 1$ & $-$ \\
& $E_{2u}$ &$\hat p_z(\hat p_x\pm i\hat p_y)^2$	&$\cY^{\pm 2}_3(\p)$& $\pm 2$ & $-$ \\
\hline
\end{tabular}
\caption{Representative orbital basis functions, expressed in the chiral basis, for the point groups $D_{4h}$ and $D_{6h}$.\label{table-Ereps}}
\end{table}

For even-parity, spin-singlet pairing the Nambu-matrix order parameter has the form, $\widehat\Delta(\vp)=\left(\Delta(\vp)\,\tp+\right.$ $\left.\Delta^*(\vp)\,\tm\right)\otimes(i\sigma_y)$, where $\tpm=\left(\tx\pm i\ty\right)/2$ and $\{\tone,\tx,\ty,\tz\}$ are $2\times 2$ matrices spanning particle-hole (Nambu) space. The spin-singlet correlations are represented by the Pauli matrix $i\sigma_y$, which is anti-symmetric under exchange. The orbital order parameter, $\Delta(\vp)$, is symmetric under exchange implying $\Delta(-\vp) = +\Delta(\vp)$. The general form of the orbital order parameter is spanned by the two-dimensional space of $\point{E}{1(2)g}$ basis functions. For $\point{E}{1g}$ the chiral basis $\{\cY_{\nu}(\vp)|\nu = \pm 1\}$ can be constructed from the 2D vector representation: $\cY_{\nu}(\vp)=\cY_{zx}(\vp)+i\nu\cY_{zy}(\vp)$ $=\sin(p_z a_z)(\hat{p}_x+i\nu\hat{p}_y)=\sin(p_z a_z)\,e^{i\nu\phi_{\vp}}$, where the latter two forms correspond to $\point{E}{1g}$ basis functions defined on a cylindrical Fermi surface with $\phi_{\vp}$ corresponding to the azimuthal angle of $\vp$.
Note that chiral $\point{E}{1g}$ pairing also breaks reflection symmetry in the plane normal to the chiral axis, and has a line of nodes in the energy gap for momenta in the plane $p_z = 0$. Thus, $\point{E}{1g}$ pairing is not realized in 2D, but is defined on a 3D Fermi surface.

For chiral $\point{E}{2g}$ pairing the basis functions can be defined as $\cY_{\nu}(\vp)=\cY_{x^2-y^2}(\vp)+i\,\sgn(\nu)\cY_{xy}(\vp)$ $=(\hat{p}_x+i\sgn(\nu)\hat{p}_y)^{|\nu|}=e^{i\nu\phi_{\vp}}$, with $\nu=\pm 2$. The the latter two forms correspond to $\point{E}{2g}$ pairing defined on a cylindrical Fermi surface.
Note that the chiral ground state for $\point{E}{2g}$ also breaks reflection symmetry in one or more planes containing the chiral axis, $\hat\vell=\hat\vz$, but, in contrast to $\point{E}{1g}$, preserves reflection symmetry in the plane normal to the chiral axis. Thus, a fully-gapped chiral ground state is possible in 2D, as well as a 3D Fermi surface that is open in the $p_z$ direction.
For a 3D Fermi surface that is closed in the $p_z$ direction, the chiral $\point{E}{2g}$ ground state has topologically protected nodal points of $\Delta(\vp)$ at $\vp_{\pm} =\pm p_f\hat\vz$, and a corresponding spectrum of massless chiral Fermions in the bulk phase~\cite{gos15}.

For odd-parity, ESP triplet states the Nambu-matrix order parameter takes the form, $\widehat\Delta(\vp)=\left(\Delta(\vp)\,\tp-\right.$ $\left.\Delta^*(\vp)\,\tm\right)\otimes(\sigma_x)$, where we have chosen the ESP state with $\hat\vd=\hat\vz$~\footnote{Results for heat and charge transport in zero field do not depend on the choice for the direction of $\hat{\vd}$.}.
The ESP triplet-correlations are represented by the symmetric Pauli matrix $\sigma_x$, and the odd-parity orbital order parameter, $\Delta(\vp)$, which is necessarily anti-symmetric under exchange, i.e. $\Delta(-\vp) = -\Delta(\vp)$. 
For $\point{E}{1u}$ pairing the chiral basis $\{\cY_{\nu}(\vp)|\nu = \pm 1\}$ is constructed from the odd-parity 2D vector representation: $\cY_{\nu}(\vp)=\cY_{x}(\vp)+i\nu\cY_{y}(\vp)$ $=(\hat{p}_x+i\nu\hat{p}_y) = e^{i\nu\phi_{\vp}}$, the latter two forms correspond to $\point{E}{1u}$ basis functions defined on a cylindrical Fermi surface with $\phi_{\vp}$ corresponding to the azimuthal angle of $\vp$. 
In contrast to $\point{E}{1g}$, the $\point{E}{1u}$ chiral ground states are fully gapped in 2D, and in 3D for an open Fermi surface in the $p_z$ direction. For chiral $\point{E}{2u}$ pairing the basis functions are constructed from those of $\point{E}{2g}$ by multiplying by odd-parity function of $p_z$. Thus, $\cY_{\nu}(\vp)=\cY_{z(x^2-y^2)}(\vp)+i\sgn(\nu)\cY_{z(xy)}(\vp)$ $=\sin(p_z a_z)(\hat{p}_x+i\sgn(\nu)\hat{p}_y)^{2}=\sin(p_z a_z)\,e^{i\nu\phi_{\vp}}$, with $\nu=\pm 2$. These chiral states correspond to the $\point{E}{2u}$ pairing model for the B-phase of \upt. 

\subsection{2D Chiral Superconductors\label{sec-2D}}

Here we consider the fully gapped $\point{E}{1u}$ and $\point{E}{2g}$ chiral ground states defined on a 2D cylindrical Fermi surface. These two cases illustrate nearly all of the key physical phenomena responsible for anomalous thermal and electrical transport mediated by non-magnetic impurity scattering in chiral superconductors.
At low temperatures, thermally excited quasiparticles and phonons are dilute, therefore quasiparticles interact predominantly with quenched defects. For randomly distributed impurities, the self-energy is given by $\hat\Sigma_{\imp}(\p;\e)=n_{\imp}\hat t(\p,\p;\e)$, where $n_{\imp}$ is the mean impurity density and $\hat t(\p,\p;\e)$ is the forward scattering limit of the single-impurity $t$-matrix in the superconducting state. This $t$-matrix can be expressed in terms of the normal-state $t$-matrix, and the latter can be expressed in terms of scattering phase shifts for normal-state quasiparticle-impurity scattering,
\begin{subequations}
\label{eq:tmat_eq}
\begin{align}
\hat t^{R,A}(\p',\p;\e)
&=\hat t_N^{R,A}(\p',\p) 
+ N_f\left\langle\hat t_N^{R,A}(\p',\p'')\left[\hat g^{R,A}(\p'';\e)-\hat g_N^{R,A}\right]
 \hat t^{R,A}(\p'',\p;\e)\right\rangle_{\p''}
\,,
\\
\hat t^K(\p',\p;\e)
&=N_f\left\langle\hat t^R(\p',\p'';\e)
\hat g^K(\p'';\e)\hat t^A(\p'',\p;\e)\right\rangle_{\p''}
\,,
\end{align}
\end{subequations}
where $N_f$ is the normal-state density of states per spin at the Fermi surface and $\langle\dots\rangle_{\p}$ represents averaging over the Fermi-surface---for an isotropic 2D Fermi surface, $\langle\dots\rangle_{\p}=\int_0^{2\pi}d\phi_{\p}/(2\pi)(\dots)$. The superscripts denote three types of quasiclassical propagators: retarded ($R$), advanced ($A$) and Keldysh ($K$). In deriving Eq.~\eqref{eq:tmat_eq}, the bare electron-impurity interaction is eliminated in favor of the normal-state propagator, $\hat g_N=-\pi g_N\hat\tau_3$ with $g_N^{R}=(g_N^{A})^*=i$, and the normal-state t-matrix,
\begin{equation}
\hat t_N(\p',\p)=
\frac{-1}{\pi N_f}
\sum_{m=-\infty}^{+\infty} \frac{e^{im(\phi-\phi')}}{\cot\delta_{m}-g_N\hat\tau_3}
\end{equation}
with $\delta_m$ the scattering phase shift in the $m^\text{th}$ cylindrical harmonic~\footnote{The summation over $m$ is truncated as a defect with characteristic radius $R$ leads to rapidly decaying phase shifts for $|m|\gtrsim k_fR$.}.
Here and in the following, the directions ($\p,\p',\p'',\dots$) on the Fermi surface and their corresponding azimuth angles ($\phi,\phi',\phi'',\dots$) are used interchangeably.

The mean field order parameter for unitary chiral states can be expressed in the following form, ${\widehat{\Delta}}_{S}(\vp)={\widehat{U}}_{S}\,\widehat{\Delta}(\vp)\,{\widehat{U}}_S^\dagger$, where ${\widehat{U}}_{S}$ is the unitary matrix for singlet ($S=0$) or triplet ($S=1$) pairing,
\be
\widehat{U}_{0} =
\bmat 
i\sigma_y &  0
\\
0	  & 1
\emat
\,,
\quad
\widehat{U}_{1} =
\bmat
\hat\vd\cdot i\vsigma\sigma_y &	0
\\
0	& 	1
\emat
\,,
\label{eq-spin_structure}
\ee
in which case $\widehat\Delta(\vp)$ reduces to
\be\label{eq-Mean-Field_OP-2D}
\widehat{\Delta}(\vp) 
=
\Delta\,e^{i\nu\phi_{\vp}\tz}\,(i\ty)
=
\bmat
0	&  \Delta\,e^{i\nu\phi_{\vp}}
\\
-\Delta\,e^{-i\nu\phi_{\vp}} & 0
\emat
\,,
\ee
for both $S=0$ and $S=1$. 
Thus, in the absence of external magnetic fields, magnetic impurities or spin-dependent perturbations, the spin structure of the order parameter can be transformed away by a unitary transformation, and as previously noted the quasiparticle excitation spectrum is doubly degenerate with respect to the quasiparticle spin.

This representation of the mean-field order parameter extends to the off-diagonal components of the impurity self energy. In Eq.~\eqref{eq-Mean-Field_OP-2D} we chose $\Delta$ to be real. In this gauge the off-diagonal impurity self-energies reduce to 
\begin{align}
\widehat{\Delta}_{\text{imp}}^{R,A}(\vp;\varepsilon)
=
\Delta_{\text{imp}}^{R,A}(\varepsilon)\,e^{i\nu\phi_{\vp}\tz}(i\ty)
\,,
\label{eq:gap}
\end{align}
with the gauge condition, $\Delta_{\text{imp}}^{R,A}(\varepsilon)=\Delta_{\text{imp}}^{R,A}(-\varepsilon)^*$.
The Nambu-matrix impurity self energy can then be expressed in terms of three functions 
\be\label{eq:defo_D}
\widehat\Sig_{\text{imp}}^{R,A}(\vp;\varepsilon)
=
D^{R,A}_{\text{imp}}(\varepsilon)\tone 
+ 
\Sig_{\text{imp}}^{R,A}(\varepsilon)\tz 
+ 
\Delta^{R,A}_{\text{imp}}(\varepsilon)\,e^{i\nu\phi_{\vp}\tz}(i\ty)
\,.
\ee
The term proportional to the unit Nambu matrix, $D^{R,A}_{\text{imp}}(\varepsilon)\tone$, drops out of Eq.~\eqref{eq-QC-RA_equilibrium} for the equilibrium propagators, $\widehat{g}^{R,A}$, and thus plays no role in determining the equilibrium properties of the superconductor. However, the unit-matrix term does contribute to the linear response of the superconductor, e.g.\  the a.c.\ conductivity~\cite{gra95}. 

The diagonal term proportional to $\tz$ can be combined with the excitation energy and expressed as
\be
\tilde{\varepsilon}^{R,A}(\varepsilon) = \varepsilon - \Sig_{\text{imp}}^{R,A}(\varepsilon)
\,,
\ee
and similarly the impurity renormalized off-diagonal self energy is given by
\be
\tilde{\Delta}^{R,A}(\varepsilon) = \Delta + \Delta_{\text{imp}}^{R,A}(\varepsilon)
\ee
Thus, for any of the chiral, unitary states described by Eq.~\eqref{eq:gap}, the equilibrium propagators that satisfy the transport equation and normalization condition, Eqs.~\eqref{eq-QC-RA_equilibrium} and \eqref{eq-normalization-equilibrium}, are given by
\begin{align}
\widehat{g}^{R,A}(\vp;\varepsilon) & = -\pi\frac{\tilde\varepsilon^{R,A}\tz 
- 
\tilde\Delta^{R,A}\,e^{i\nu\phi_{\vp}\tz}(i\ty)}{\sqrt{(\tilde\Delta^{R,A})^2-(\tilde{\varepsilon}^{R,A})^2}}
\\
\label{eq:defo_g_f}
&\equiv - \pi [g^{R,A}(\varepsilon)\tz+f^{R,A}(\varepsilon)\,e^{i\nu\phi_{\vp}\tz}(i\ty)]
\,.
\end{align}
Note that the functions $g^{R,A}$ and $f^{R,A}$ satisfy the symmetry relations: $g^{A}=(g^{R})^*$ and $f^{A}=(f^{R})^*$. In equilibrium, the Keldysh propagator is determined by the spectral functions for quasiparticles and Cooper pairs, and the thermal distribution of excitations,
\be
\widehat{g}^{K}(\vp;\varepsilon)
= 
\left[
\widehat{g}^{R}(\vp;\varepsilon)
- 
\widehat{g}^{A}(\vp;\varepsilon)
\right]\tanh\frac{\varepsilon}{2T}
\,.
\ee

\subsubsection*{Gap Equation: mean-field order parameter}

The pairing interaction combined with the off-diagonal component of the Keldysh propagator determines the mean-field pairing self-energy for any of the unitary chiral states is given by the ``gap equation'',
\be
\Delta(\vp) 
= 
\int d\vp'\,\lambda(\vp,\vp')\,\fint \frac{d\varepsilon'}{4\pi i}\,f^{K}(\vp';\varepsilon')
\ee
where the pairing interaction in any of the two-dimensional E-reps defined on a cylindrical Fermi surface has the form
\begin{align}
\lambda(\vp,\vp') 
= 
\lambda_{|\nu|}\,
\left(e^{-i\nu\phi_{\vp}}\,e^{+i\nu\phi_{\vp'}}
+ e^{+i\nu\phi_{\vp}}\,e^{-i\nu\phi_{\vp'}}
\right)
=
2\lambda_{|\nu|}\,\cos[\nu(\phi_{\vp}-\phi_{\vp'})]
\,.    
\end{align}
Thus, projecting out the amplitude of the chiral mean-field order parameter we obtain the gap equation,
\be\label{eq-gap}
\Delta
= \lambda_{|\nu|}\,
\fint\frac{d\varepsilon}{4\pi i}\,(\pi)[f^R(\e)-f^A(\e)]\tanh\frac{\varepsilon}{2T}
\,.
\ee
In practice the pairing interaction strength $\lambda_{|\nu|}$ is eliminated in favor of the critical temperature. 

The equilibrium retarded and advanced propagators are given by 
\begin{align}
\hat{g}^{R,A}(\p;\e)
= 
-\pi\frac{\tilde\e^{R,A}(\e)\hat\tau_3-\tilde\Delta^{R,A}(\e)e^{i\hat\tau_3\nu\phi}(i\hat\tau_2)}{C^
{R,A}(\e)}
=-\pi\left[g(\e)\hat\tau_3+f(\e)e^{i\hat\tau_3\nu\phi}(i\hat\tau_2)\right]
\,,
\end{align}
where $g=\tilde\e/C$, $f=-\tilde\Delta/C$ and $C=[{\tilde\Delta}(\e)^2-\tilde\e(\e)^2]^{1/2}$~\footnote{Hereafter the retarded (R) and advanced (A) superscripts are not shown for $g(\varepsilon)$, $f(\varepsilon)$ etc., but are implied.}. The equilibrium spectrum is renormalized by interactions with impurities, i.e., $\tilde\e=\e-\Sigma_{\imp}$ and ${\tilde\Delta}=\Delta+\Delta_{\imp}$, where $\Delta$ is the mean-field excitation gap from Eq.~\eqref{eq-gap}, and $\Sigma_{\imp}$ and $\Delta_{\imp}$ are the diagonal and off-diagonal terms in the impurity self energy, Eq.~\eqref{eq:defo_D}~\footnote{Despite its absence from spectral renormalization, $D^{R,A}(\e)$ encodes particle-hole asymmetry, e.g.\ the difference in scattering lifetimes for particles and holes, which could have implications for transport properties~\cite{Arfi:1989}, especially in thermoelectric responses~\cite{Lofwander:2004}.}.
The self-energy is obtained from the equilibrium $\hat t$ matrix,
\begin{align}
\hat t(\p',\p;\e)
&=
\frac{-1}{\pi N_f}
\begin{pmatrix}
t(\p',\p;\e)&a(\p',\p;\e)\\-\ul a(\p',\p;\e)&\ul t(\p',\p;\e)
\end{pmatrix}
=
\frac{-1}{\pi N_f}\sum_m e^{im(\phi-\phi')}
\begin{pmatrix}
t_m(\e)&e^{i\nu\phi}a_{-m}(\e)\\-e^{-i\nu\phi}\ul a_{m}(\e)&\ul t_{-m}(\e)
\end{pmatrix}.
\end{align}
Upon solving Eqs.~\eqref{eq:tmat_eq}, we obtain
\begin{subequations}
\label{eq:equilibrium_tmatrix}
\begin{align}
\genfrac{\{}{\}}{0pt}{0}{t_{m}(\e)}{\ul t_{m}(\e)}
&=
\frac{\sin\delta_m[\cos\delta_{m+\nu}\pm g(\e)\sin\delta_{m+\nu}]}{
\cos(\delta_m-\delta_{m+\nu})\mp g(\e)\sin(\delta_m-\delta_{m+\nu})}
\\%
\genfrac{\{}{\}}{0pt}{0}{a_{m}(\e)}{\ul a_{m}(\e)}
&=
\frac{f(\e)\sin\delta_m\sin\delta_{m-\nu}}{
\cos(\delta_m-\delta_{m-\nu})\mp g(\e)\sin(\delta_m-\delta_{m-\nu})}.
\label{eq:equilibrium_amatrix}
\end{align}
\end{subequations}
The diagonal terms $t_m$ and $\ul t_m$ are the amplitudes for quasi-particles and quasi-holes scattering off an impurity with relative angular momentum $m$.
The off-diagonal terms, $a_{m}$ and $\ul a_{m}$, are the amplitudes for branch conversion scattering in which a quasi-particle (quasi-hole) scatters off an impurity and also converts to a quasi-hole (quasi-particle). The branch conversion process is accompanied by the creation (destruction) of a Cooper pair. In a chiral superconductor the Cooper pairs have angular momentum $\nu\hbar$, and thus branch conversion scattering requires a corresponding change in the angular momentum of the scattered excitation, e.g.\ $m\rightarrow m-\nu$ for an incident quasi-particle scattering with relative angular momentum $m\hbar$ converting to a quasi-hole and a Cooper pair of angular momentum $\nu\hbar$.
Thus, for branch conversion scattering to occur the quasiparticle-impurity potential must support amplitudes, $a_m$,  and thus non-vanishing phase shifts $\delta_m$ with $|m|\in\{0,1,\ldots,|\nu|\}$, as can be seen from Eq.~\eqref{eq:equilibrium_amatrix}.
A direct consequence is that isotropic impurity scattering from point-like impurities does not support branch conversion scattering in chiral superconductors since the incoming and outgoing scattering states are limited to the s-wave ($m=0$) scattering channel.
As we show in what follows, the ionic radius of the impurity and branch conversion scattering are central in determining the magnitude and temperature dependence of anomalous Hall transport in chiral superconductors.
Finally we note that the propagators, t-matrix and self-energies must be computed self-consistently with the gap equation, Eq.~\eqref{eq-gap}. 
In Sec.~\ref{sec-2D_results} we summarize results for thermal transport in 2D fully gapped chiral superconductors with a random distribution of finite size impurities.

\subsection{3D Chiral Superconductors}\label{sec-3D}

Here we consider chiral superconductors in 3D belonging to the two-dimensional E-representations of the tetragonal ($\point{D}{4h}$) and hexagonal ($\point{D}{6h}$) point groups, both even- and odd-parity E$_1$ and E$_2$ representations. These groups describe the discrete point symmetries of \sro and \upt, respectively. See Table \ref{table-Ereps}.

\subsubsection{Symmetries of the order parameter}

The mean-field pairing self-energy, after factoring the spin-structure using the unitary transformation in Eq.~\eqref{eq-spin_structure}, has the structure,
\begin{equation}
\widehat\Delta^{R,A}(\p) 
= 
\begin{pmatrix}
0	&	\Delta(\p)	\\	-\Delta(\p)^*	&	0
\end{pmatrix}
\,.
\end{equation}
The weak-coupling mean-field order parameter, ``gap function'', is independent of energy and related to the equilibrium Keldysh pair propagator by the BCS gap equation,
\begin{equation}
\Delta(\p)
=
\fint\frac{d\e}{4\pi i}
\la\lambda(\p,\p')f^K(\p';\e)\ra_{\p'}
\,,
\end{equation}
where $\lambda(\p,\p')$ provides the pairing interaction and the energy integral is over the bandwidth of attraction, $\varepsilon_c$, $\p$ and $\p'$ are the directions of the relative momentum of pairs of Fermions with zero total momentum, and $\la\dots\ra_\p$ is an average over the Fermi surface.
The pairing interaction respects the maximal symmetry of the point group and can be expressed as a sum over bi-linear products of basis functions of the irreducible representations analogous to Eq.~\eqref{eq-Cooper-channel_irreps}. We assume the irreducible representation, $\Gamma$, with the most attractive pairing interaction, $\lambda_{\Gamma}$, dominates, in which case we can ignore the sub-dominant pairing channels,
\begin{equation}
\lambda(\p,\p') = \lambda_\Gamma\sum_{\nu}^{\text{dim}_\Gamma}\,
\eta_{\Gamma,\nu}(\p)\eta_{\Gamma,\nu}(\p')^*
\,,
\end{equation}
where the interaction amplitude $\lambda_\Gamma$ determines the critical temperature and $\eta_{\Gamma,\nu}(\p)$ denotes the basis functions for the irreducible representation, $\Gamma$, of the relevant point group. Table~\ref{table-Ereps} summarizes the basis functions, expressed in chiral basis, for the point groups $D_{4h}$ and $D_{6h}$, and which are relevant for \sro\ and the heavy-fermion compound \upt, respectively~\footnote{A comprehensive set of tables of basis functions for the tetragonal, hexagonal and cubic point groups is provided in Ref.~\cite{mineev99}.}.
For a chiral superconductor the order parameter is proportional to one of the chiral basis functions, e.g.\ $\Delta(\p)\propto\eta_{\Gamma,\nu}(\p)$ for a chiral ground state belonging to the irrep $\Gamma$ with winding number $\nu$. For the analysis to follow it is sufficient to consider pairing of states near a 3D spherical Fermi surface, in which case the mean-field pairing self-energy is proportional to the spherical harmonic, i.e., $\eta_{\Gamma,\nu}(\p)=Y_l^{\nu}(\p)$, where $l$ is the orbital angular momentum corresponding to the irrep $\Gamma$,
\begin{equation}\label{eq:gapmatrix}
\Delta(\p)=\Delta\,{\widetilde Y}_l^{\nu}(\p)
\,,
\end{equation}
where $\Delta$ is the maximum value of the order parameter, and the normalized spherical harmonics are related to the standard spherical harmonics $Y_l^m(\p)$ via ${\widetilde Y}_l^m(\p)=Y_l^m(\p)/\max_{\p}|Y_l^m(\p)|$.

\subsubsection{Impurity self-energy}

In the low temperature limit quasiparticle scattering from thermally excited quasiparticles and phonons is negligible compared to scattering off the random impurity potential. For a homogeneous uncorrelated random distribution of impurities the corresponding self-energy is a product of the mean impurity density $n_\imp$ and the forward scattering limit of the single impurity $t$-matrix,
\begin{equation}
\widehat\Sig^{R,A}_\imp(\p;\e)=n_\imp\widehat t^{\,R,A}(\p,\p;\e)
\,,
\end{equation}
where the $t$-matrices are obtained as a solution of the integral equation,
\begin{equation}\label{eq:t-matrix_R}
\widehat t^{\,R,A}(\p',\p)
=
\widehat t_N^{\,R,A}(\p',\p)
+
N_f
\la\widehat t_N^{\,R,A}(\p',\p'')
\left[\g^{\,R,A}(\p'')-\g_N^{\,R,A}\right]
\widehat t^{\,R,A}(\p'',\p)\ra_{\p''}
\,.
\end{equation}
The Keldysh component of the $t$-matrix then given by
\begin{equation}\label{eq:t-matrix_K}
\hspace{-2mm}
\widehat t^{\,K}(\p',\p)=N_f\la\widehat t^{\,R}(\p',\p'')\g^{\,K}(\p'')\widehat t^{\,A}(\p'',\p)\ra_{\p''}
\,.
\end{equation}
Note that $\widehat t^A$ can be obtained from the symmetry relation, $\widehat t^{A}(\p',\p)=\tz\widehat t^{\,R}(\p,\p')^\dagger\tz$.
In Eqs.~\eqref{eq:t-matrix_R} and \eqref{eq:t-matrix_K} we eliminated the electron-impurity matrix element, $\widehat u$, in favor of the normal-state quasiparticle propagator, $\g_N=-\pi g_N\tz$ with $g_N^{R}=(g_N^{A})^*=i$, and the normal-state $t$-matrix,
\begin{equation}
\widehat t_N(\p',\p)
=\frac{-1}{\pi N_f}\sum_{l\ge0}\,\sum_{m=-l}^{+l}\frac{Y^{m}_l(\p')Y^m_l(\p)^*}{\cot\delta_{l}+\g_N/\pi}	
\,,
\end{equation}
where $\delta_l$ is the phase shift in the relative angular momentum channel, $l$. The normalization of the spherical harmonics is given by
\begin{equation}
\langle Y^m_l(\p)^*Y^{m'}_{l'}(\p)\rangle_\p=\delta_{mm'}\delta_{ll'}
\,.
\end{equation}
An important feature of scattering theory by central force potentials, in this case the quasiparticle-impurity potential, is that the characteristic range $R$ of the potential leads to phase shifts $\delta_l$ that decay rapidly to zero for $l\gtrsim k_f R$, effectively truncating the summations over $m$ and $l$.

\subsubsection{Equilibrium Properties}

Below we present the framework for determining the self-consistent equilibrium propagators. To highlight the effects of chiral phase winding we consider systems that are gauge-rotation invariant, i.e. invariant under a rotation around the chiral axis combined with a specific element of the $\point{U(1)}{}$. 
As a result the diagonal equilibrium propagator depends on $\p$ only through the polar angle $\theta_p$ measured from the chiral axis, $g(\p;\e)=g(\theta_p;\e)$. The azimuth angle $\phi_p$ appears only in the phase factor of the pair propagator, i.e. $f(\p;\e)=e^{i\nu\phi_{p}}f(\theta_p;\e)$, where $\nu$ is winding number of the chiral order parameter. Thus we parametrize the propagators as
\begin{equation}
\g^{R,A}(\p;\e)
=
-\pi
\left[g^{R,A}(\theta_p;\e)\tz
+
f^{R,A}(\theta_p;\e)e^{i\nu\tz\phi_p}(i\ty)
\right]
\,,
\end{equation}
where the quasiparticle and pair propagators read 
\begin{equation}
g^{R,A}=\frac{\tilde\e^{R,A}(\theta;\e)}{C^{R,A}(\theta;\e)}	
\quad\text{and}\quad
f^{R,A}=-\frac{\tilde\Delta^{R,A}(\theta;\e)}{C^{R,A}(\theta;\e)}
\,,
\end{equation}
with
\begin{equation}
\label{eq:def_C}
C^{R,A}(\theta;\e)
=
\sqrt{\tilde\Delta^{R,A}(\theta;\e)^2-\tilde\e^{R,A}(\theta;\e)^2}.
\end{equation}
The equilibrium Keldysh propagator is determined by $\widehat g^{R,A}$ and the Fermi distribution function,
\begin{equation}
\g^K=\left(\g^R-\g^A\right)\tanh\frac{\e}{2T}
\,.
\end{equation}
Note that the retarded and advanced propagators are related by the symmetry relation, $\g^{A}=\tz(\g^{R})^\dagger\tz$.
The excitation energy and order parameter are renormalized by impurity scattering through the impurity self-energies, $\Sigma_\imp$ and $\Delta_\imp$, via
\begin{equation}\label{eq:spectral_renorm}
\begin{aligned}
\tilde\e^{R,A}(\theta;\e)	
&=
\e-\Sigma_\imp^{R,A}(\theta;\e) \\
\tilde\Delta^{R,A}(\theta;\e)
&=
\Delta\,\widetilde\Theta^\nu_l(\theta)
+
{\Delta}_\imp^{R,A}(\theta;\e),
\end{aligned}
\end{equation}
where $\widetilde\Theta^\nu_l(\theta)$ is the polar-angle dependence of the normalized spherical harmonics. The functions $\Sigma_\imp$ and $\Delta_\imp$ are defined such that
\begin{equation}
\label{eq:def_D}
\widehat\Sigma_\imp^{R}(\p;\e)
=D^{R}(\theta_p;\e)\widehat1	
+\Sigma_\imp^{R}(\theta_p;\e)\tz
+\Delta_\imp^{R}(\theta_p;\e)e^{i\tz\nu\phi_p}(i\ty)	
\,.
\end{equation}
The function $D(\e)$ encodes the asymmetry in scattering rates for particles and holes, which has implications for the thermoelectric response of chiral superconductors~\cite{Lofwander:2004}.

Since the scattering potential is rotationally invariant we can expand the $t$-matrix equation \eqref{eq:t-matrix_R} into a set of decoupled equations for each cylindrical harmonic channel. Thus, we parametrize the $t$-matrix as follows
\begin{equation}
\label{eq:cylindrical_tmat}
\widehat t^{\,R}(\p',\p)		
=
\sum_{m}\,e^{im(\phi_p-\phi_{p'})}
\left[
\widehat t_{m}^{\,R}(\theta_{p'},\theta_p)
+
e^{i\nu\tz\phi_{p'}}
\,
\widehat a^{R}_{m}(\theta_{p'},\theta_p)
\right]
\,,
\end{equation}
where the diagonal part of the $t$-matrix is given by
\begin{equation}
\widehat t_{m}^{\,R}(\theta_{p'},\theta_p)
=
\begin{pmatrix}
t_{m}^{R}(\theta_{p'},\theta_p) & 0 \\
0 & \ul t_{m}^{R}(\theta_{p'},\theta_p)
\end{pmatrix}
\,,
\end{equation}
and the off-diagonal part is given by
\begin{equation}
\widehat a_{m}^{R}(\theta_{p'},\theta_p)
=
\begin{pmatrix}
0 & a^{R}_{m}(\theta_{p'},\theta_p) \\
-\ul a^{R}_{m}(\theta_{p'},\theta_p) & 0
\end{pmatrix}
\,.
\end{equation}
Thus, by factoring out the dependence on the azimuth angle as shown in Eq.~\eqref{eq:t-matrix_R}, we obtain integral equations for the cylindrical harmonics of the $t$-matrix,
\begin{eqnarray}
\widehat t_{m}^{\,R}(\theta_{p'},\theta_p)
&=&
\widehat t_{N,m}^{\,R}(\theta_{p'},\theta_p)
+
\la	
\widehat t_{N,m}^{\,R}(\theta_{p'},\theta_{p''})
[
(g^R(\theta_{p''})-g^R_N)\tz\widehat t_{m}^{\,R}(\theta_{p''},\theta_p)
+
f^R(\theta_{p''})(i\ty)\,
\widehat a^{R}_{m}(\theta_{p''},\theta_p)
]
\ra_{p''}
\qquad
\label{eq:cylindrical_tmat_eq}
\\
\widehat a^{R}_{m}(\theta_{p'},\theta_p)
&=&
\la	
\widehat t_{N,m-\nu\tz}^{\,R}(\theta_{p'},\theta_{p''})
\left[
f^R(\theta_{p''})(i\ty)
\widehat t_{m}^{\,R}(\theta_{p''},\theta_p)
+
(g^R(\theta_{p''})-g^R_N)\tz\,
\widehat a^{R}_{m}(\theta_{p''},\theta_p)
\right]
\ra_{p''}
\,,
\label{eq:cylindrical_amat_eq}
\end{eqnarray}
where $\widehat t_{N,m-\nu\tz}=\Diag(t_{N,m-\nu},\ul t_{N,m+\nu})$. The off-diagonal $t$-matrix $\widehat a_{m}$ describes Andreev scattering in which an incoming particle branch converts into an outgoing hole and vice versa.
This process relies on the creation and destruction of a Cooper pair. For chiral pairing, the conservation of angular momentum implies that the cylindrical harmonics of incoming and outgoing scattering states must differ by the orbital angular momentum quantum number of a Cooper pair, hence the phase winding factor in front of $\widehat a_{m}$ in Eq.~\eqref{eq:cylindrical_tmat}. Since branch conversion scattering requires two distinct angular momentum channels, branch conversion scattering is absent for point-like impurities which support only s-wave ($l=0$) scattering.
Equations.~\eqref{eq:cylindrical_tmat_eq} and \eqref{eq:cylindrical_amat_eq} are solved for the 3D Fermi surface by expanding in the spherical harmonic basis functions, $\Theta^m_l(\theta)$.
For cylindrical Fermi surfaces the dependence on $\theta$ drops out and the $t$-matrices are obtained by matrix inversion as described in Sec.~\ref{sec-2D}.

\section{Linear Response Theory}

For small departures from equilibrium driven by a small temperature bias between different edges of the superconductor the heat current is proportional to the temperature gradient, 
\be
{\bf j}^{(q)}=-\tnsr\kappa\cdot\boldsymbol\nabla T
\,,
\ee 
where $\tnsr\kappa$ is the thermal conductivity tensor which is constrained by the chiral symmetry of the ground state. To obtain these transport coefficients, $\kappa_{ij}$, we determine the self-consistent, equilibrium propagators \emph{and} their first-order non-equilibrium corrections to linear order in $\boldsymbol\nabla T$.
The equilibrium propagators encode information about the bound and unbound quasiparticle pair spectrum, and are key inputs to the determination of the linear-response functions. 
The heat current is computed from the non-equilibrium Keldysh propagator in Eq.~\eqref{heat-current}.

Here we consider the linear response functions for a static and homogeneous thermal gradient. For convenience we separate the Keldysh response into a spectral and anomalous part. The anomalous response encodes information about the non-equilibrium distribution function and is defined by,
\be
\delta\widehat x^a(\e)=\delta\widehat x^K(\e)-\tanh(\e/2T)\left[\delta\widehat x^R(\e)-\delta\widehat x^A(\e)\right],
\ee
where $\widehat x$ stands for the propagator ($\widehat x\to\widehat g$) or self-energy ($\widehat x\to\widehat\Sig$). We focus on the anomalous functions because the spectral response functions, $\delta\widehat x^{R,A}$, do not contribute to the thermal conductivity tensor in linear response theory to leading order in the quasiclassical expansion parameters~\cite{gra96a}.
For a uniform thermal gradient the anomalous propagator is obtained from the solution of the linearized transport equations (see Ref.~\cite{gra96a} for the general solution),
\begin{equation}
\delta\g^a=-\frac{C^{a}_+\g^{R}_\text{eq}/\pi+D^{a}_-}{(C^{a}_+)^2+(D^{a}_-)^2}
\left[
(\g^R_\text{eq}-\g^A_\text{eq})(i{\bf v}_\p\cdot\nabla\Phi)
+(\g^R_\text{eq}\delta\widehat\Sig^a-\delta\widehat\Sig^a\g^A_\text{eq})
\right],
\label{eq:dg^a}
\end{equation}
where $\nabla\Phi=\nabla\tanh[\e/2T({\mathbf r})]$ is the gradient of the local equilibrium distribution function. We added the subscript ``eq'' to denote the equilibrium propagators. We also adopt the shorthand notation,
\begin{equation}
C^a_+=C^R+C^A
\quad\text{and}\quad
D^{a}_-=D^R-D^A,
\end{equation}
with $C$ and $D$ defined in Eqs.~\eqref{eq:def_C} and \eqref{eq:def_D}, respectively.
It is also efficient to express the response functions as column vectors whose elements correspond to those of their corresponding matrices in particle-hole space,
\begin{eqnarray}
\ket{\delta g(\p;\e)}&=\left(\delta g(\p;\e),\delta\ul g(\p;\e),\delta f(\p;\e),\delta\ul f(\p;\e)\right)^T						
\,,
\label{eq:response_vector-g}
\\
\ket{\delta\Sig(\p;\e)}&=\left(\delta\epsilon(\p;\e),\delta\ul\epsilon(\p;\e),\delta\Delta(\p;\e),\delta\ul\Delta(\p;\e)\right)^T
\,.
\label{eq:response_vector-Sigma}
\end{eqnarray}
The expression for the anomalous propagator (Eq.~\ref{eq:dg^a}) can then be recast as
\begin{equation}
\label{eq:linear_response1}
\ket{\delta g^a(\p;\e)}=\mathbb L^a(\p;\e)
\Bigg[
\ket{\psi^a(\p;\e)}+\ket{\delta\sigma^a(\p;\e)}
\Bigg]
\,,
\end{equation}
where the static thermal gradient leads to the perturbation,
\begin{equation}
\ket{\psi(\p;\e)}
=
i{\bf v}_\p\cdot\nabla\Phi\,(1,1,\cdot,\cdot)^T
\equiv
\psi(\p;\e)\,(1,1,\cdot,\cdot)^T
\,.
\end{equation} 
The linear-response matrix $\mathbb L(\p;\e)$ is defined in terms of the equilibrium propagators,
\begin{multline}
{\mathbb L}^a(\e)= 
-
{\mathcal C}^a
\begin{pmatrix}
1+|g|^2	&	-|f|^2		&	-g^Rf^A	&	-f^Rg^A	\\
-|f|^2	&	1+|g|^2 	&	f^Rg^A	&	g^Rf^A	\\
g^Rf^A	&	-f^Rg^A		&	1-|g|^2	&	-|f|^2	\\ 
f^Rg^A	&	-g^Rf^A		&	-|f|^2	&	1-|g|^2
\end{pmatrix}
-
{\mathcal D}^a
\begin{pmatrix}
g^R-g^A		& \cdot		&	f^A		&	-f^R	\\
\cdot		& -g^R+g^A	&	-f^R		&	f^A	\\
-f^A		& f^R		&	g^R+g^A		&	\cdot	\\
f^R		& -f^A		&	\cdot		&	-g^R-g^A
\end{pmatrix}
\,,
\end{multline}
where $|g|^2=g^Rg^A$, $|f|^2=f^Rf^A$ and
\begin{equation}
\{{\mathcal C}^a(\e),{\mathcal D}^a(\e)\}
=
\frac{1}{2}\frac{\{\Re\, C^R(\e),i\,\Im\, D^R(\e)\}}
                {[\Re\, C^R(\e)]^2-[\Im\, D^R(\e)]^2}
\,.
\end{equation}

\subsection{Self-Energy - Vertex Corrections\label{sec:imp_vertex}} 

The {\it r.h.s.} of Eq.~\eqref{eq:linear_response1} consists of two terms. The first is the contribution that is explicitly proportional to the external field, $\vv_p\cdot\grad T$. This term contributes only to the longitudinal thermal conductivity. Indeed the anomalous Hall conductivity arises solely from the non-equilibrium self-energy term.
The self-energy corrections are the \emph{vertex corrections} in the field-theoretical formulation based on Kubo response theory. These terms describe the response to perturbations by long-wavelength collective excitations of the interacting Fermi system~\cite{rai94b}. In the context of the linear response theory developed for disordered chiral superconductors, the vertex corrections resulting from interactions of Bogoliubov quasiparticles with static impurities are obtained from the linear response corrections 
to the equilibrium $t$-matrix Eqs.~\eqref{eq:t-matrix_R} and \eqref{eq:t-matrix_K} obtained from the the first-order non-equilibrium corrections to the full $t$-matrix Eqs.~\eqref{eq-tmatrix-RA} and \eqref{eq-tmatrix-K}. 
For the anomalous self-energy expressed in Nambu matrix form,
\begin{equation}
\label{eq:vertex_correction_original}
\delta\widehat\Sig^{a}_\imp(\p)
=
n_\imp N_f
\la\widehat t^{\,R}_\text{eq}(\p,\p')\delta\g^{\,a}(\p')\widehat t^{\,A}_\text{eq}(\p',\p)\ra_{\p'}
\,,
\end{equation}
can be recast in column vector form as defined by Eq.~\eqref{eq:response_vector-Sigma}, 
\begin{equation}\label{eq:imp_ver_corr}
|\delta\Sig^a(\p)\rangle
=
-\frac{n_\imp}{\pi N_f}
\Big\langle\mathbb T^a(\p,\p')\ket{\delta g^a(\p')}\Big\rangle_{\p'}\,,
\end{equation}
where the impurity vertex-correction operator is given by
\begin{equation}
\mathbb T^a(\p,\p')
=
\begin{pmatrix}
t^R t^A		&	-a^R\ul a^A	&	-t^R\ul a^A		&	-a^R t^A	\\
-\ul a^R a^A	&	\ul t^R\ul t^A	&	-\ul a^R \ul t^A	&	-\ul t^R a^A	\\
t^R a^A		&	a^R \ul t^A	&	t^R\ul t^A		&	-a^R a^A	\\
\ul a^R t^A	&	\ul t^R \ul a^A	&	-\ul a^R\ul a^A		&	\ul t^R t^A	
\end{pmatrix}
\,.
\end{equation}
The retarded [advanced] $t$-matrix elements are evaluated at $(\p,\p';\e)$ [$(\p',\p;\e)$], and the equilibrium $t$-matrix elements, $t$ and $a$, are defined such that
\begin{equation}
\widehat t^{\,R,A}_\text{eq}(\p',\p)
=
\begin{pmatrix*}[r]
t^{R,A}(\p',\p)		&	a^{R,A}(\p',\p)\\
-\ul a^{R,A}(\p',\p)	&	\ul t^{R,A}(\p',\p)
\end{pmatrix*}
\,.
\end{equation}


In general the mean-field pairing self-energy also contributes a vertex correction (i.e. $\delta\widehat\Sig=\delta\widehat\Delta+\delta\widehat\Sig_\imp$). These terms play a central role in collective mode response of the condensate, however, in the present context they contribute only to the retarded and advanced self-energies. The vertex correction contributing to anomalous heat transport arises only from the impurity-induced self-energy.


For point-like impurities, the vertex correction, and thus the anomalous Hall current, vanishes in all but chiral p-wave states. This can be shown by noting that for isotropic impurity scattering the vertex correction from Eq.~\eqref{eq:vertex_correction_original}, $\delta\widehat\Sig(\e)\propto\langle\delta\widehat g(\p;\e)\rangle_{\p}$, is obtained from the isotropic components of the anomalous propagator. 
The diagonal components of the equilibrium propagators are isotropic, and thus their contribution to the linear response function involves momentum dependence only from the perturbation, $\psi(\p)\propto\vv_{\hat\vp;\e}\cdot\grad\Phi$. This p-wave term vanishes when averaged over the Fermi-surface, and as a result does not contribute a vertex correction. On the other hand, the off-diagonal components of the equilibrium propagators acquire the phase factor $e^{\pm i\nu\phi}$, reflecting the angular momentum of the chiral Cooper pairs.
The linear response functions from these terms include a phase factor $e^{i(\pm\nu+1)\phi}$, which contributes a vertex correction only when $|\nu|=1$, i.e. for chiral p-wave pairing. This is a non-universal result specific to point-like impurities. For finite-size impurities vertex corrections and thus anomalous Hall effects result for chiral superconductors with $|\nu| > 1$ with results varying with the ionic radius of the impurity. 

\subsection{Cylindrical harmonic decomposition for 2D chiral superconductors}

For chiral superconductors with cylindrically symmetric Fermi surfaces and pairing interactions we can parametrize the non-equilibrium corrections to the propagators and self-energies in terms of cylindrical harmonics,
\be
\begin{aligned}
\delta\widehat g(\p;\e) 
&=
-\pi\sum_n e^{in\phi} 
\begin{pmatrix}
\delta g_n(\e)			&	e^{i\nu\phi}\delta f_n(\e) \\
-e^{-i\nu\phi}\delta\ul f_n(\e)	&	\delta\ul g_n(\e)
\end{pmatrix}
\,,
\\
\delta\widehat\Sig(\p;\e) 
&=
\sum_n e^{in\phi} 
\begin{pmatrix}	
\delta\epsilon_n(\e)			&	e^{i\nu\phi}\delta\Delta_n(\e)	\\
-e^{-i\nu\phi}\delta\ul\Delta_n(\e)	&	\delta\ul\epsilon_n(\e)	
\end{pmatrix}\
\,.
\end{aligned}
\ee
The response in different cylindrical harmonic channels can decoupled such that Eq.~\eqref{eq:linear_response1} reduces to
\be
\label{eq:lineartransporteq}
|\delta g^a_n(\e)\rangle	=	\mathbb L^a(\e)
\Big[	|\psi^a_n(\e)\rangle	+	|\delta\Sig^a_n(\e)\rangle\Big]
\,,
\ee
where 
$|\delta g_n\rangle=(\delta g_n,\delta\ul g_n,\delta f_n,\delta\ul f_n)^T$, 
$|\delta\Sigma_n\rangle=(\delta\epsilon_n,\delta\ul\epsilon_n,\delta\Delta_n,\delta{\ul\Delta}_n)^T$, 
and the temperature gradient along the $x$ axis generates the perturbation
\be
\label{eq:sigmaPhi}
|\psi_n^a(\e)\rangle	=	\delta_{|n|,1}	\psi_1^a(\e)	(1,1,\cdot,\cdot)^T
\ee
with $\psi_1^a(\e)=-i\,\frac{\e v_f}{4T^2}\sech^2(\frac{\e}{2T})\,\partial_xT$.
The impurity self-energy correction from Eq.~\eqref{eq:imp_ver_corr} becomes
\be
|\delta\Sig^a_{n}(\e)\rangle	=	-\frac{n_{\imp}}{\pi N_f}
\mathbb T_{n}^a(\e)	|\delta g^a_n(\e)\rangle
\,,
\label{eq:lineartmatrixeq}
\ee
where the vertex-correction operators are given by 
\be\mathbb T^a_{n}(\e)
=
\langle\langle\mathbb{Y}_n(\p)^*\mathbb 
T^a(\p,\p';\e)\mathbb{Y}_n(\p')\rangle_{\p}\rangle_{\p'},
\ee 
with $\mathbb{Y}_{n}(\p)=e^{in\phi}\Diag(1,1,e^{i\nu\phi},e^{-i\nu\phi})$.
Substituting Eq.~\eqref{eq:lineartmatrixeq} into Eq.~\eqref{eq:lineartransporteq} results in a linear matrix equation for the cylindrical harmonics of the anomalous response.

\subsection{Spherical harmonic decomposition for 3D chiral superconductors}

To exploit the axial symmetry of the Fermi surface and chiral symmetry of the order parameter, we write the anomalous propagator ($\delta x\to\delta  g$) and self-energy ($\delta x\to\delta\Sig$) as a sum of spherical harmonic components
\begin{equation}
|\delta x(\p;\e)\rangle	=\sum_l\sum_m{\mathbb Y}^m_l(\p)|\delta x_{l;m}(\e)\rangle		
\end{equation}
with
\begin{equation}
{\mathbb Y}^l_m(\p)\equiv\text{Diag}\Big[Y^{m}_l(\p),Y^{m}_l(\p),Y^{m+\nu}_l(\p),Y^{m-\nu}_l(\p)\Big]
\,.
\end{equation}
The spherical harmonic components are then given by
\begin{equation}
|\delta x_{l;m}(\e)\rangle 
=
\Big\langle
{\mathbb Y}^m_l(\p)^*\left|\delta x(\p;\e)\right\rangle\Big\rangle_{\hat\vp}
\,.
\end{equation}
The anomalous response in Eq.~\eqref{eq:linear_response1} can now be expressed in terms of solutions for each cylindrical harmonic component,
\begin{equation}\label{eq:g^a_spherical_harmonics}
|\delta g^a_{l;m}(\e)\rangle
=
\sum_{l'}\mathbb L^a_{ll';m}(\e)
\Big[
|\psi^a_{l';m}(\e)\rangle+|\delta\Sig^a_{l';m}(\e)\rangle
\Big]\,,
\end{equation}
where the perturbation is
\begin{equation}
\ket{\psi^a_{m;l}(\e)}
=
\la Y^m_l(\p)^*(i{\bf v}_\p\cdot\nabla\Phi)\ra_{\hat\vp}
(1,1,\cdot,\cdot)^T	
\end{equation}
and the linear response matrix is given by
\begin{equation}
\mathbb L^a_{ll';m}(\e)
=
\la\mathbb Y^m_l(\p)^*\mathbb L^a(\p;\e)\mathbb Y^m_{l'}(\p)\ra_{\hat\vp}
\,.
\end{equation}
Similarly the vertex correction, Eq.~\eqref{eq:imp_ver_corr}, is recast as
\begin{equation}\label{eq:s^a_spherical_harmonics}
\ket{\delta\Sigma_{l;m}^a}
=
-\frac{n_\imp}{\pi N_f}
\sum_{l'}\mathbb T^a_{ll';m}\ket{\delta g^a_{l';m}},
\end{equation}
where 
\begin{equation}
\mathbb T^a_{ll';m}
=
\la\la\mathbb{Y}^m_l(\p)^*\mathbb T^a(\p,\p')\mathbb{Y}^m_{l'}(\p')\ra_{\hat\vp}\ra_{\hat\vp'}
\,.
\end{equation}
Finally we use Eq.~\eqref{eq:s^a_spherical_harmonics} to eliminate the self-energy term from Eq.~\eqref{eq:g^a_spherical_harmonics}, yielding
\begin{equation}\label{eq:linear-response-eq}
\sum_{l'}\left[\mathbb 1\delta_{ll'}+\frac{n_\imp}{\pi N_f}
\sum_{k}\mathbb L^a_{lk;m}\mathbb T^a_{kl';m}\right]
\ket{\delta g^a_{l';m}}
=
\sum_{l'}\mathbb L^a_{ll';m}\ket{\psi^a_{l';m}}
\,.
\end{equation}
This equation is solved by matrix inversion.

\begin{figure}
\centering
\includegraphics[width=0.75\linewidth]{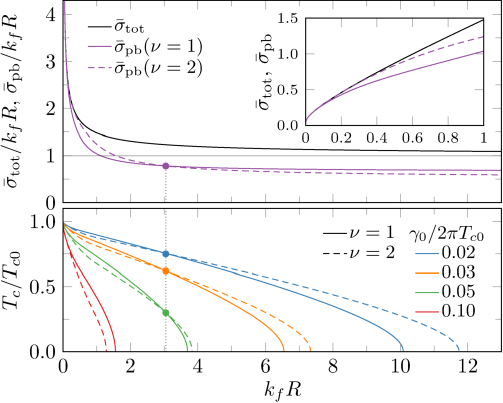}
\caption{Impurity cross sections and critical temperature versus hard-disc radius for chiral states: $\nu=1$ (solid) and $\nu=2$ (dashed). For hard-disc radius $k_fR\approx3.05$, pair-breaking cross sections and critical temperature of the two states coincide (filled circles). {\it Top:} Total cross section (black), transport cross section (solid purple) and pair-breaking cross section (purple). {\it Bottom:} Critical temperature for various impurity densities: $\gamma_0/2\pi T_{c_0} \equiv n_\imp\xi_0/k_f$ (see legend). 
Figure reproduced from Ref.~\cite{wave:20} with permission of the American Physical Society (APS) and the authors.}
\label{fig:Tc_harddisc}
\end{figure}

\section{Results for 2D Chiral superconductors}\label{sec-2D_results}

To quantify the effects of finite-size impurities, we consider hard-disc scattering for which the scattering phase shifts are given by $\tan\delta_m = J_{|m|}(k_fR)/N_{|m|}(k_fR)$~\cite{Lapidus:1986}, where $R$ is the hard-disc radius and, $J_m(z)$ and $N_m(z)$ are Bessel functions of the first and second kind, respectively. Results presented in this section were reported in Ref.~\cite{wave:20}. They are included here to highlight the effects of disorder on fully gapped topological chiral superconductors and to compare with new results for 3D nodal chiral superconductors. We start with the effects of impurities on the equilibrium properties and the sub-gap excitation spectrum.

\subsection{Suppression of \texorpdfstring{$T_c$}{Tc} and Pair-breaking} 

For temperatures approaching the critical temperature, $T_c$, from below temperature the order parameter approaches zero continuously at the second order transition. The resulting linearized gap equation yields the transition temperature in terms of the pairing interaction, $\lambda$, bandwidth of attraction (``cutoff''), $\varepsilon_c$, and the pair-breaking effect of quasiparticle-impurity scattering. The pairing interaction and cutoff can be eliminated in favor of the clean-limit transition temperature, $T_{c_0}$, with the result being a transcendental equation for the suppression of $T_c$ from quasiparticle scattering off the random distribution of impurities.
The resulting critical temperature is given by~\footnote{Similar results were derived for the suppression of $T_c$ by non-magnetic disorder in p-wave superconductors and superfluid \He\ in aerogel~\cite{lar65,thu98}.} 
\be\label{eq-Tc-2D_disorder}
\ln\frac{T_{c_0}}{T_c}
=
\Psi\left(\frac12	
+
\frac{1}{2}\frac{\xi_0\sigma_\text{pb}n_\imp}{T_c/T_{c_0}}\right)
-
\Psi\left(\frac12\right)
\,,
\ee 
where $\Psi(x)$ is the digamma function, $T_{c_0}$ is the critical temperature and $\xi_0=v_f/2\pi T_{c_0}$ is the coherence length in the clean limit. The effects of pair-breaking by impurity scattering is determined by the pair-breaking cross section
\be\label{eq:sigma_pb}
\sigma_\text{pb}=\frac{2}{k_f}\sum_m\sin^2(\delta_m-\delta_{m+\nu})
\,,
\ee
for a chiral order parameter with winding number $\nu$. For s-wave pairing ($\nu=0$), $\sigma_\text{pb}=0$ and consequently $T_c=T_{c_0}$ as expected from Anderson's theorem~\cite{and59}. In Fig.~\ref{fig:Tc_harddisc}, we see that $\sigma_\text{pb}$ is generally \emph{different} from the total cross section $\sigma_\text{tot}=(4/k_f)\sum_m\sin^2\delta_m$. The two cross sections approach one another only in the point-like impurity limit $k_fR\ll 1$. Furthermore, $\sigma_\text{pb}$ and $T_c$ both depend on the ionic radius and the winding number. A feature of the hard-disk scattering model is that $\sigma_\text{pb}$ for $\nu=2$ and $\nu=1$ cross at $k_f R\approx3.05$. For radii smaller (larger) than this value, pair breaking is stronger and $T_c$ is lower for $\nu=2$ ($\nu=1$).

\begin{figure}
\centering
\includegraphics[width=0.75\linewidth]{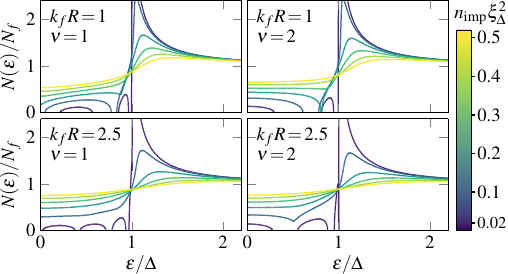}
\caption{
Density of states for chiral order $\nu=1$ (left) and $\nu=2$ (right), various impurity densities normalized by $\xi_\Delta^2=(\pi N_f\Delta)^{-1}$ (see color bar), and impurity radii: $k_fR=1$ (top) and $2.5$ (bottom).
Figure reproduced from Ref.~\cite{wave:20} with permission of the APS and the authors.
}
\label{fig:dos}
\end{figure}

\subsection{Density of States}

The quasiparticle spectrum, $N(\e)=N_f\Im\,g^R(\e)$, also depends sensitively on the winding number, $\nu$, as shown in Fig.~\ref{fig:dos}. Note the existence of multiple sub-gap impurity bound states, which are broadened into bands with increasing impurity density. These states are generated by the combination of potential scattering by impurities and multiple Andreev scattering by the chiral order parameter. As a result, the number of bound states and their sub-gap energies are determined by not only the impurity potential, e.g.\ the ionic radius, but also the winding number $\nu$. The impurity-induced sub-gap spectrum has important implications for all quasiparticle transport processes.
In the low-temperature limit, $T\ll|\Delta|$, the thermal conductivity is dominated by excitations at energies well below the clean limit gap edge. Diffusion within the lowest energy band of sub-gap states near the Fermi level determines the low temperature heat current as we discuss below.

\subsection{Thermal Conductivity and the Anomalous Thermal Hall Effect} 

\begin{figure}
\centering
\includegraphics[width=0.8\linewidth]{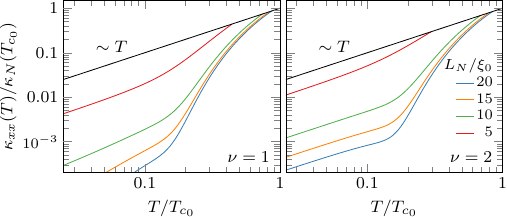}
\caption{Longitudinal thermal conductivity versus temperature for chiral superconductors with $\nu=1$ (left) and $\nu=2$ (right), impurity radius $k_f R=1$, and normal-state transport lengths listed in the legend. The normal-state thermal conductivity is shown in black.
Figure reproduced from Ref.~\cite{wave:20} with permission of the APS and the authors.%
}
\label{fig:kappaxx_log}
\end{figure}

\begin{figure}
\centering
\includegraphics[width=0.75\linewidth]{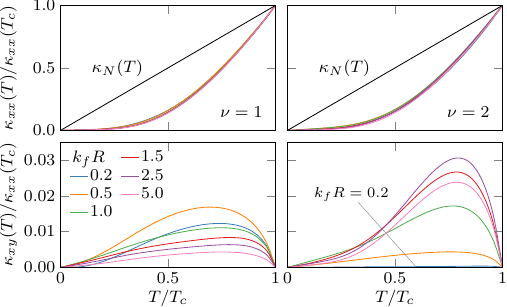}
\caption{Longitudinal (top) and transverse (bottom) thermal conductivity vs temperature for chiral order $\nu=1$ (left) and $\nu=2$ (right), normal-state transport length $L_N/\xi_0=7.5$, and various impurity radii (see legend). Normal-state thermal conductivity shown in black.
Figure reproduced from Ref.~\cite{wave:20} with permission of the APS and the authors.%
}
\label{fig:kappa_xy_Tc}
\end{figure}

In normal metals the thermal conductivity is limited by the transport mean free path for quasiparticles scattering off the random distribution of impurities, $\kappa_N=(\pi^2/3)N_fv_fL_N T$, where $L_N=1/(\sigma_\text{tr}\,n_\imp)$ is determined by the transport cross-section~\footnote{N.B. The transport and pair-breaking cross sections are different except for $|\nu|=1$, c.f. Eq.~\eqref{eq:sigma_pb}.}. 
\be
\sigma_\text{tr}=(2/k_f)\sum_m\sin^2({\delta_{m}-\delta_{m+1}})
\,.
\ee
In the superconducting state the thermal conductivity depends on both the mean impurity density as well as the impurity cross-section via,
\be
\label{eq:heat_conductivity}
\kappa_{\genfrac{\{}{\}}{0pt}{2}{xx}{xy}}(T)
=
N_fv_f\int d\e\left(\frac{\e}{2T}\sech\frac{\e}{2T}\right)^2L_{\genfrac{\{}{\}}{0pt}{2}{xx}{xy}}(\e),
\ee
where we define the thermal transport lengths for the longitudinal and transverse currents by
\be
L_{xx}(\e)
\equiv
\Re\frac{v_f\delta g^a_{1}(\e)}{-2\psi^a_{1}(\e)}
\quad\text{and}\quad
L_{xy}(\e)
\equiv
\Im\frac{v_f\delta g^a_{1}(\e)}{-2\psi^a_{1}(\e)}
\,.
\ee

Figure~\ref{fig:kappaxx_log} shows the temperature dependence of longitudinal thermal conductivity for fully gapped chiral superconductors with $\nu=1,2$. Note that the presence of impurities generally enhances the low-temperature thermal conductivity through the formation of sub-gap states, but the enhancement depends on winding number of the chiral order. For impurities with $k_f R=1$ note that a band of Andreev bound states with a finite density of states at $\e=0$ develops for a chiral order parameter with $\nu=2$, but not for $\nu=1$ as shown in Fig.~\ref{fig:dos}. This is because the state with $\nu=2$ has more phase space for scattering on the Fermi surface with a nearly perfect sign change that leads to maximal pairbreaking (i.e. scattering with $\delta\vartheta\approx \pm\pi/2$) compared to the state with $\nu=1$ (scattering with $\delta\vartheta\approx \pi$).
Thus, for $\nu=2$ a gapless, diffusive, ``metallic'' band results in a low-temperature thermal conductivity which is linear in temperature as $T\rightarrow 0$, i.e. $\kappa_{xx}(T\to 0)\propto T$. We also note that for $\nu=1$, such behavior only occurs for sufficiently large impurity densities where the impurity bands broaden to close the gap at $\e=0$.

Figure~\ref{fig:kappa_xy_Tc} illustrates perhaps the most pronounced effects of finite-size impurities on transport properties. Although the longitudinal conductivity is relatively insensitive to the impurity size or the winding number, the Hall conductivity depends strongly on both $R$ and $\nu$. For point-like impurities with radii smaller than the Fermi wavelength, $k_f R\lesssim 1$, the thermal Hall conductivity is finite for $\nu=1$, but is dramatically suppressed for chiral states with $|\nu|> 1$, as is clear in the comparison between $\nu=1$ and $\nu=2$ for $k_fR=0.2$ shown in the lower two panels of Fig.~\ref{fig:kappa_xy_Tc}.
This supports our previous argument that Hall currents vanish for point-like impurities, i.e. $k_f R\ll 1$, for all chiral winding numbers except $|\nu|=1$. Also note that as we increase the radius of the impurities such that $k_f R \gtrsim 1$, the Hall conductivity for $\nu=2$ increases dramatically and can be substantially larger than that for $\nu=1$. Furthermore, for a fixed normal-state transport mean free path, the Hall conductivity exhibits a non-monotonic dependence on impurity size, reaching maximum at an intermediate radius. Thus, the details of the impurity potential, and thus the sub-gap spectrum, are of crucial importance for a quantitative understanding of anomalous Hall effects in chiral superconductors.

\begin{figure}
\centering
\includegraphics[width=0.75\linewidth]{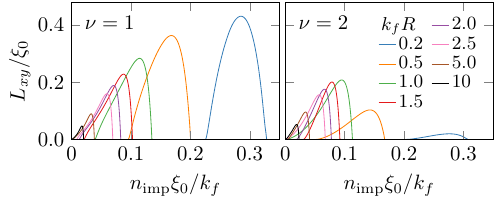}
\caption{Thermal Hall transport length at $\e=0$ for chiral order $\nu=1$ (left) and $\nu=2$ (right), and varying impurity radii (see legend). Low-temperature transport requires quasiparticles states at $\e=0$, formed only with adequate impurity density. But high impurity density destroys superconductivity and thus rules out any anomalous Hall effects.%
}
\label{fig:L_xy}
\end{figure}

It is also instructive to compare the low-temperature limit of thermal Hall transport originating from the bulk topology in the form of chiral edge states with the bulk thermal Hall conductance from the random distribution of impurities embedded in the bulk of the superconductor. For chiral p-wave pairing the edge-state contribution to the thermal Hall conductance $K_{xy}^\text{edge}/T=\pi k_B^2/6\hbar$ is universal~\cite{rea00,Senthil:1999,Sumiyoshi:2013}.
By contrast the bulk impurity contribution to the low-temperature thermal Hall conductance can be expressed as $K_{xy}^\text{bulk}=k_fL_{xy}^{\e=0}\times K_{xy}^\text{edge}$ [see Eq.~\eqref{eq:heat_conductivity}], where $L_{xy}^{\e=0}$ is the effective transport length, which is non-universal and depends on the impurity density and scattering cross-section. Furthermore, $L_{xy}^{\e=0}$ is finite in a finite range of impurity density for which there is a finite density of states at $\e=0$, but not so disordered as to destroy superconductivity, as shown in Fig.~\ref{fig:L_xy}. 
At sufficiently low impurity density the spectrum is gapped at $\e=0$ and so the edge contribution, which is linear in $T$ can dominate at sufficiently low temperatures. While above a critical impurity density superconductivity is destroyed and with it the Hall transport. However, over a significant range of impurity density both the edge and bulk impurity contributions are present for all $T < T_c$.

To compare the edge and bulk contributions to $K_{xy}$ when both are present we consider typical values of the coherence length to Fermi wavelength, $k_f\xi_0$, and the relative impurity size, $k_fR$. For example, taking $k_f\xi_0=100$, $\nu=1$ and $k_fR=0.5$ we find $K_{xy}^\text{bulk}\approx 35 K_{xy}^\text{edge}$ at the value of $n_\imp$ that maximizes $L_{xy}$ as shown in Fig.~\ref{fig:L_xy}.
In general we find that the bulk contribution to the anomalous thermal Hall conductivity is generally dominant over most of the temperature range.

\section{Results for Chiral superconductors in 3D}

We have extended the analysis for chiral states in 2D to chiral states defined on closed 3D Fermi surfaces which often include symmetry enforced line and point nodes of the excitation gap. The results reported here include anomalous thermal Hall effects in candidates for 3D chiral superconductors belonging to tetragonal and hexagonal crystalline point groups, particularly the perovskite \sro\ and the heavy-fermion superconductor \upt.
To investigate the effects of ionic radius and the dependence on the ionic cross-section, we use the hard-sphere impurity potential for which the scattering phase shifts are analytically given in terms of the hard-sphere radius, $R$, and the Fermi wavevector~\cite{messiah58a},
\begin{equation}
\tan\delta_l = \frac{j_{l}(k_fR)}{n_{l}(k_fR)},
\end{equation}
where $j_l(z)$ and $n_l(z)$ are spherical Bessel functions of the first and second kind, respectively~\cite{abramowitz70}.

\subsection{Critical Temperature}

\begin{figure}
\centering
\includegraphics[width=0.75\linewidth]{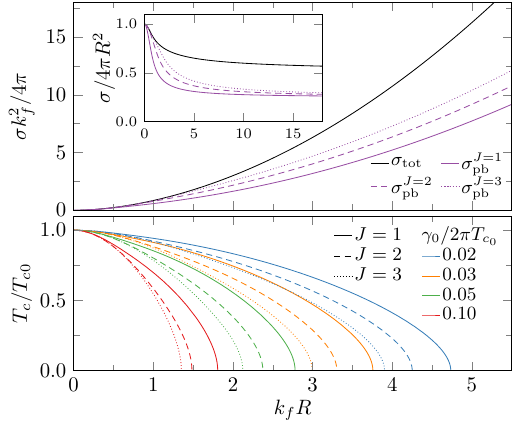}
\caption{{\it Top:} Total cross section (black) and pair-breaking cross section (purple) versus hard-sphere radius for pairing with total angular momentum $J=1$ (solid) $J=2$ (dashed) and $J=3$ (dotted). Note that the transport cross section is $\sigma_\text{tr}=\sigma_\text{pb}^{J=1}$.
{\it Inset:} Scattering cross sections in units of $4\pi R^2$.
{\it Bottom:} Critical temperature versus impurity radius for the same pairing states shown in the top panel for various impurity densities shown in the legend where $\gamma_0=n_\text{imp}/\pi N_f$.
}
\label{fig:Tc}
\end{figure}

For 3D chiral superconductors we obtain a result of the same form as Eq.~\eqref{eq-Tc-2D_disorder} for the suppression of $T_c$ by disorder, but with a pair-breaking cross-section appropriate for scattering of a 3D Fermi surface with finite-size impurities in 3D,
\begin{equation}\label{eq:Tc-formula}
\ln\frac{T_{c_0}}{T_c}
=
\Psi\left(\frac12+\frac{n_\imp\xi_0\sigma_\text{pb}}{2T_c/T_{c_0}}\right)-\Psi\left(\frac12\right)
\,,
\end{equation}
where the pair-breaking cross section is given by, 
\begin{equation}\label{eq:sigmapb}
\sigma_\text{pb}
=
\frac{4\pi}{k_f^2}\frac12\sum_{ll'}\sin^2(\delta_l-\delta_{l'})
\frac{(2l+1)(2l'+1)}{2J+1}
\langle l,0;l',0|J,0\rangle^2
\,,
\end{equation}
with $\la l,m;l',m'|L,M\ra$ are Clebsch-Gordan coefficients and $J=|\nu|$ is the Cooper pair angular momentum quantum number. For s-wave pairing ($J=0$) the Clebsch-Gordan coefficient vanishes unless $l=l'$, and thus $\sigma_\text{pb}=0$ and $T_c=T_{c_0}$, consistent with Anderson's theorem \cite{and59}. Note also that the pair-breaking cross-section is in general different from both the total cross-section and transport cross section, which are defined by
\begin{align}
\label{eq:sigmatot}
\sigma_\text{tot} &=\frac{4\pi}{k_f^2}\sum_{l}(2l+1)\sin^2\delta_l \\
\sigma_\text{tr}  &=\frac{4\pi}{k_f^2}\sum_{l}(l+1)\sin^2(\delta_{l+1}-\delta_l)=\sigma_\text{pb}^{J=1}
\,,
\label{eq:sigmatr}
\end{align}
which determine the quasiparticle scattering lifetime and transport mean-free path, respectively. For point-like impurities all of the above cross sections coincide except for pairing in the s-wave channel, in which case $\sigma_\text{tot}=\sigma_\text{tr}$, but $\sigma_\text{pb}=0$. 

In the limit $k_f R \ll 1$ the total cross-section and pair-breaking cross-section both approach $\sigma_\text{tot}=\sigma_\text{pb}=4\pi R^2$, i.e. four times the geometric cross section of the hard sphere impurity.
However, for $k_f R\gtrsim 1$ the pair-breaking cross section is typically smaller than the total cross section as shown in Fig.~\ref{fig:Tc}. In the limit $k_f R\gg 1$ $\sigma_\text{tr}$ and $\sigma_\text{pb}^{J}$ approach the geometric limit, $\pi R^2$.
However, in general the pair-breaking cross section is dependent on the topology of the order parameter. Chiral states with higher angular momentum are subject to stronger pair-breaking effects as we show for hard-sphere impurities. In the lower panel of Fig.~\ref{fig:Tc} we show the pair-breaking effects of impurity size and concentration on the critical temperature $T_c$ as described by Eqs.~\eqref{eq:Tc-formula} and~\eqref{eq:sigmapb}.

\subsection{Quasiparticle Spectrum}

\begin{figure*}
\centering
\includegraphics[width=\linewidth]{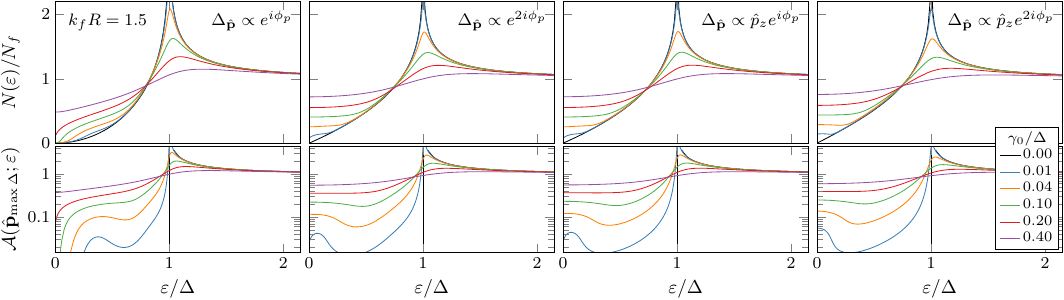}
\caption{The quasiparticle density of states $N(\e)$ (top) and spectral function along the direction of gap maximum $\mathcal{A}(\hat{\mathbf p};\e)$ (bottom) are shown for hard-sphere impurities with radius, $k_fR=1.5$, for impurity densities shown in the legend where $\gamma_0=n_\imp/\pi N_f$.
The four pairing states correspond to the chiral ground states of the irreducible representations $E_{1u}$, $E_{2g}$, $E_{1g}$ and $E_{2u}$ (left to right) of the hexagonal point group $D_{6h}$.
Scattering resonances appears as impurity-induced sub-gap bands which depend on the topology of the order parameter and the structure of the scattering potential.
}
\label{fig:DoS1.5}
\end{figure*}

\begin{figure*}
\centering
\includegraphics[width=\linewidth]{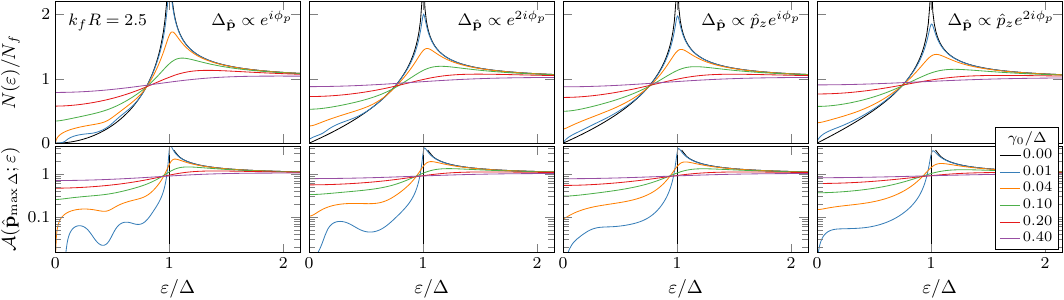}
\caption{Same description as that in Figure~\ref{fig:DoS1.5}, but for $k_fR=2.5$.
}
\label{fig:DoS2.5}
\end{figure*}

Central to the interplay between chiral symmetry, topology and disorder is the impact of impurity scattering on pair-breaking and the resulting sub-gap quasiparticle spectrum. Distinct from fully gapped 2D topological states, 3D chiral ground states support symmetry protected nodes of the order parameter which leads to quasiparticle states over the entire energy range from the maximum gap on the Fermi surface down to the Fermi energy.
The quasiparticle spectral function defines the angle-resolved quasiparticle density of states is determined by the retarded diagonal propagator, 
\begin{equation}
\mathcal A(\p;\e)=\Im\,g^R(\p;\e)
\,.
\end{equation}
The local density of states is the Fermi-surface average of the spectral function,
\begin{equation}
N(\e)=N_f\la\mathcal A(\p;\e)\ra_{\hat\vp}
\,,
\end{equation}
where $N_f$ is the normal-state density of states at the Fermi level.
Figures~\ref{fig:DoS1.5} and \ref{fig:DoS2.5} show the effects of impurity induced scattering on the quasiparticle spectrum. The coherence peak at the maximum gap edge is broadened by impurity scattering. The spectral weight is redistributed to sub-gap energies by the formation of sub-gap resonances. The formation of sub-gap impurity bands is clearly visible in the spectral function for positions on the Fermi surface corresponding to the maximum gap as shown in the bottom panel of Figs.~\ref{fig:DoS1.5} and \ref{fig:DoS2.5}.
These resonances correspond to Andreev bound states that hybridize with continuum states near nodal regions of the order parameter (c.f. Ref.~\cite{she16}).
Impurity-induced sub-gap states are formed by multiple Andreev scattering from the combined potential scattering and branch-conversion scattering by the phase-winding of the order parameter on the Fermi surface. The spectrum depends on the structure of the scattering potential as well as the topological winding number of the order parameter. 
These impurity-induced sub-gap states play a central role in determining the magnitude and temperature dependence of the anomalous thermal Hall conductivity because these states couple to the chiral condensate is the source of broken time-reversal and mirror symmetries.


\subsection{Thermal Conductivity Tensor for Chiral Superconductors} 

The heat current density in Eq.~\eqref{heat-current} for chiral ground states with embedded impurity disorder reduces to 
\begin{equation}
j^{(q)}_{\genfrac{\{}{\}}{0pt}{2}{x}{y}}	
=\frac{v_fN_f}{\sqrt6}\genfrac{\{}{\}}{0pt}{0}{\Im}{\Re}\int d\e \,\e\,\delta g^a_{1;1}(\e)
\,,
\end{equation}
where $\delta g^a_{l;m}(\e)$ is the spherical harmonic component of $g^K(\p;\e)^*=-g^K(\p;\e)$ with angular momentum quantum numbers $l,m$. In deriving this formula we used the symmetry relation $\delta g^a_{l;-m}(\e)=-(-1)^m\delta g^a_{l;m}(\e)^*$.
We also note that in linear response theory a thermal gradient does not generate a spectral response, in which case the anomalous response is equal to the Keldysh propagator, $\delta g^a = \delta g^K$.
From ${\bf j}^{(q)}=-\tnsr\kappa\cdot\nabla T$, we can express the longitudinal and transverse components of the thermal conductivity tensor as 
\begin{equation}\label{eq:heat_conductivity_3D}
\kappa_{\genfrac{\{}{\}}{0pt}{2}{xx}{xy}}	
=	
\frac43 N_fv_fT	\int \frac{d\e}{2T}
\left(\frac{\e}{2T}\sech\frac{\e}{2T}\right)^2
\,
L_{\genfrac{\{}{\}}{0pt}{2}{xx}{xy}}(\e)
\,,
\end{equation}
where the spectral resolved transport mean free paths are defined by
\begin{equation}
L_{xx}(\e)\equiv\Re\frac{v_f\delta g^a_{1;1}(\e)}{-2\psi^a_{1;1}(\e)}
\quad\text{and}\quad
L_{xy}(\e)\equiv\Im\frac{v_f\delta g^a_{1;1}(\e)}{-2\psi^a_{1;1}(\e)}
\label{eq:transport_length_defo}
\end{equation}
with $\psi^a_{1;1}(\e)=\langle Y^1_1(\p)^*(i{\bf v}_\p\cdot\nabla\Phi)\rangle_p$ and the thermal gradient is chosen to be along the $x$-axis, $\nabla\Phi=\hat{\mathbf{x}}\,\nabla_x\Phi$.
In Eq.~\eqref{eq:heat_conductivity_3D} the derivative of the Fermi distribution leads to the factor $(\e/2T)^2\sech^2(\e/2T)$ which confines the quasiparticle contribution to the heat current to excitations with $|\e|\lesssim T$.
Thus, if the transport mean free paths, $L_{xx,xy}(\e)$, vary with $\e$ on a scale $\gamma^*\gg T$, then they may be approximated by $L_{xx,xy}(\e=0)$, in which case the integration over the spectrum and thermal distribution yields,
\begin{equation}
\label{eq:heat_conductivity_zeroT}
\kappa_{\genfrac{\{}{\}}{0pt}{2}{xx}{xy}}(T) \simeq \frac{2\pi^2}{9} N_f v_f L_{\genfrac{\{}{\}}{0pt}{2}{xx}{xy}}(\e=0)\,\times T\,,\quad T\ll\gamma^*
\,.
\end{equation}
In the normal state, $\gamma^*\sim E_f\gg T$, and the above formula yields the well known result for 
the normal-state thermal conductivity with $L_{xx}(0)$ given by the transport mean-free path. In particular, in the normal state the matrices that determine the anomalous response and vertex corrections are
\begin{equation}
\begin{aligned}
\mathbb T^a_{m;ll'}  &={\mathbb 1}\delta_{ll'}\frac{k_f^2}{4\pi}(\sigma_\text{tot}-\sigma_\text{tr}) \\
\mathbb L^a_{m;ll'}  &={\mathbb 1}\delta_{ll'}\frac{-2/v_f}{\sigma_\text{tot}n_\imp}		     ,
\end{aligned}
\end{equation}
where $\sigma_\text{tot}$ and $\sigma_\text{tr}$ given by Eqs. \eqref{eq:sigmatot} and \eqref{eq:sigmatr}. Then Eq.~\eqref{eq:linear-response-eq} yields the anomalous response function,
\begin{equation}
|\delta g^a_{m;l}\rangle = \delta_{l,1}\delta_{|m|,1}
\frac{-2/v_f}{\sigma_\text{tr}n_\imp }|\psi^a_{m;1}\rangle
\,.
\end{equation}
Combining Eqs.~\eqref{eq:transport_length_defo} and \eqref{eq:heat_conductivity_zeroT} yields the normal-state thermal conductivity
\begin{eqnarray}
\kappa_{xx} &\equiv& \kappa_N = \frac{2\pi^2}{9}N_f v_f L_N\,\times T
\,,
\label{eq:thermal_conductivity_normal}
\\
\mbox{where}\quad 
L_{xx} &\equiv& L_N =\frac{1}{n_\imp\sigma_\text{tr}}
\,,
\label{eq:transport_length_normal}
\end{eqnarray}
is the transport mean free path. Furthermore, the normal state does not break time-reversal and mirror symmetries and thus $L_{xy}$ vanishes.

\begin{figure}
\centering
\includegraphics[width=0.75\linewidth]{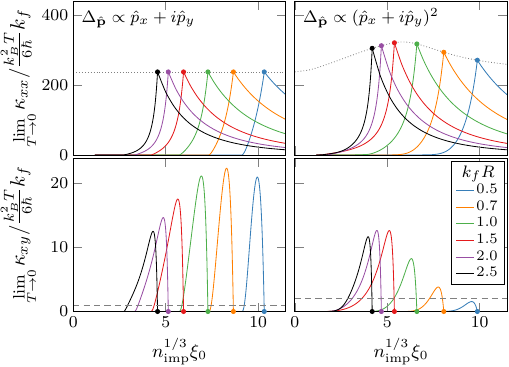}
\caption{The components of the thermal conductivity tensor, $\kappa_{xx}$ (top) and $\kappa_{xy}$ (bottom), for $T\ll\gamma^*$ as a function of impurity density and hard-sphere radii (legend) for the pairing states, $E_{1u}$ (left) and $E_{2g}$ (right) of $D_{6h}$.
The filled circles mark the critical impurity concentrations $n_\imp^c$ above which superconductivity breaks down. The dotted curves in the top panels trace the maximum value of $\kappa_{xx}$, while the dashed lines in the bottom panels show the Berry-phase contribution to the anomalous thermal Hall conductivity (Eq.~\ref{eq:kappa_surface}). All results assume $k_f\xi_0=100$.
}
\label{fig:K_zeroT_pointnode}
\end{figure}

In the chiral superconducting phase, $\gamma^*$ is a low energy scale set by the width of the impurity band at the Fermi level, $\e=0$. When it exists a metallic-like band develops which at very low temperature in the superconducting state gives rise to diffusive heat transport that is again linear in temperature for $T < \gamma^*$, now for both the longitudinal and Hall conductivities.
This regime is shown in Fig.~\ref{fig:K_zeroT_pointnode} for both components of the conductivity tensor for the pairing states $E_{1u}$ and $E_{2g}$, i.e. the states with $\Delta_{\hat{\mathbf p}}\propto e^{i\nu\phi_p}$ with $\nu=1$ and 2, respectively. For these states, low-energy excitations are located around the point nodes at $\p=\pm\hat{\bf z}$ (in the clean limit), and therefore do not contribute to low-temperature transport in the basal plane. Instead low-temperature transport relies on the impurity-induced sub-gap bands (see Figs.~\ref{fig:DoS1.5} and \ref{fig:DoS2.5}). 
The linear regime onsets for $\kappa_{xx}$ and $\kappa_{xy}$ at a threshold impurity density above which the impurity-induced resonances broaden sufficiently to generate a finite density of state at $\e=0$. The longitudinal conductivity increases with the impurity density as more states become available at the Fermi level even as the impurity scattering rate goes up. This behavior is due to the fact that the total cross section, which characterizes spectral broadening, is greater than the transport cross section as seen in Fig.~\ref{fig:Tc}. 
Above a critical impurity density, $n_\imp^c=e^{-\gamma_E}/(2\xi_0\sigma_\text{pb})$, where $\gamma_E\approx0.577$ is the Euler-Mascheroni constant, the number of available states no longer depends on impurity scattering, i.e. $N(\e=0)=N_f$. Thus, increasing the density of impurities only decreases the thermal current by reducing the transport mean free path, $L_{N}=1/n_\imp\sigma_\text{tr}$, and thus $\kappa_N=L_{N}\times k_f^2T/9$. 
%
%
\begin{figure}
\centering
\includegraphics[width=0.75\linewidth]{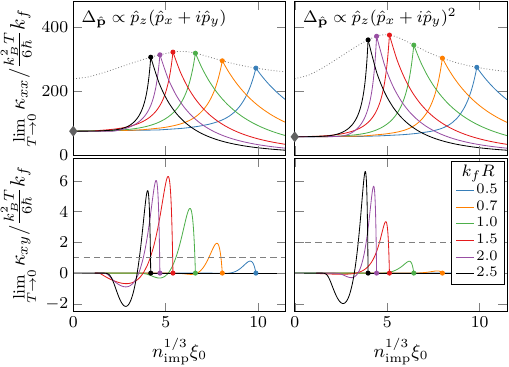}
\caption{Same plots as Fig.~\ref{fig:K_zeroT_pointnode} but for the pairing states $E_{1g}$ (left) and $E_{2u}$ (right). The diamond symbols in the top panels show the values for the ``universal limit'' for the thermal conductivity in the point-like impurity model~\cite{gra96a}.
}
\label{fig:K_zeroT_linenode}
\end{figure}
%
At $n_\imp=n_\imp^c$ the zero-temperature limit of $\kappa_{xx}/T$ is proportional to the ratio of the pair-breaking and transport cross sections, $\kappa_{xx}^c/T\propto\sigma_\text{pb}/\sigma_\text{tr}$. As a result the p-wave pairing state has a \emph{universal upper bound} for $\kappa_{xx}/T$, i.e. independent of the structure of the impurity potential, because $\sigma_\text{pb}=\sigma_\text{tr}$ for $J=1$ (Eq.~\eqref{eq:sigmatr}),
\begin{equation}
\lim_{T\to0}\frac{\kappa_{xx}}{T}\le\frac{2e^{\gamma_E}k_B^2}{9\hbar}k_f^2\xi_0
\,,\quad\text{for p-wave pairing}\,(E_{1u})
\,.
\end{equation}

The thermal Hall conductivity $\kappa_{xy}/T$ also initially increases with the impurity concentration above the lower threshold density shown in Fig.~\ref{fig:K_zeroT_pointnode}. However, $\kappa_{xy}/T$ peaks below $n_\imp^c$ as the Hall response must vanish when disorder destroys the condensate and restores time-reversal and mirror symmetries.
The thermal Hall conductivity also depends strongly on the topology of the order parameter \emph{and} the structure of impurity potential. The latter is highlighted by the comparison between the thermal Hall currents for the $E_{1u}$ and $E_{2g}$ states with decreasing impurity size. The state with $\nu=2$ is severely suppressed $k_fR<1$.
This behavior results from the suppression of branch-conversion scattering which couples impurity scattering to the order parameter of the chiral condensate. In the limit of pure s-wave impurity scattering \emph{only} the chiral states with $\Delta_{\hat{\mathbf p}}\propto e^{\pm i\phi_p}$ can support a finite Hall conductivity (see Sec.~\ref{sec:imp_vertex}).

Figure \ref{fig:K_zeroT_linenode} shows the thermal conductivity in the zero-temperature limit for the states $E_{1g}$ and $E_{2u}$, i.e. with $\Delta_{\hat{\mathbf p}}\propto \hat p_z e^{i\nu\phi_p}$ with $\nu=1$ and 2, respectively. The presence of the line node at $\hat p_z=0$ guarantees the availability of low-energy quasiparticles for transport in the basal ($x$,$y$) plane even in the clean limit. Consequently the low-temperature limit of $\kappa_{xx}/T$ does not rely solely on impurity-induced sub-gap states at the Fermi level, and is finite even for $n_\imp\to 0$ as shown in the upper panels of Fig.~\ref{fig:K_zeroT_linenode}. Indeed $\kappa_{xx}/T$ approaches universal values, identical to those obtained for point-like impurities by Ref.~\cite{gra96a}, shown as the diamond symbols in Fig.~\ref{fig:K_zeroT_linenode}.
However, the thermal Hall conductivity, $\lim_{T\to0}\kappa_{xy}/T$ does not onset at $n_\imp=0$. A finite impurity density is still essential for a non-vanishing anomalous thermal Hall conductance, $\kappa_{xy}/T$, at low temperatures. The reason is that the transverse component of the heat current is generated by by branch-conversion scattering induced by potential scattering off the distribution of impurities. 
For this process to generate a finite $\lim_{T\to0}\kappa_{xy}/T$ the sub-gap Andreev resonances must be sufficiently broadened to generate a finite density of states at the Fermi level.

\subsection{Comparison with the anomalous thermal Hall conductivity from Berry curvature}

Anomalous Hall transport in ultra-clean topological superconductors with broken time-reversal and mirror symmetries was predicted by several authors~\cite{rea00,qin11,gos15}.
In particular, anomalous Hall conductance originating from the gapless edge spectrum confined on the boundary of a topological chiral superconductor is predicted to be quantized, $\kappa_{xy}/k_{B} T=\nicefrac{\pi}{12}\,k_{B}/\hbar$. This edge contribution to the anomalous thermal Hall conductance can be computed from the Berry curvature $\Omega_{k_x,k_y}^{(n)}$ via the formula~\cite{qin11,gos15}, 
\begin{equation}
\kappa_{xy}^\text{edge} =
\frac{1}{VT}\sum_{n,\mathbf{k}}\Omega_{k_x,k_y}^{(n)}(\mathbf k)
\int_{E_{\mathbf{k},n}}^\infty d\e\,\e^2 f'(\e)
\,,
\end{equation}
where $f(\e)=\frac12(1-\tanh\frac{\e}{2T})$ denotes the Fermi-Dirac distribution, $E_{\mathbf{k},n}$, the quasiparticle spectrum with $n$ being the band index and $V$ the volume of the system. 
In the superconducting state energy eigenstates of the Bogoliubov Hamiltonian separates into two bands: above ($n=+1$) and below ($n=-1$) the Fermi level with eigenenergies $E_{\mathbf{k},\pm}=\pm\sqrt{\xi_{\mathbf k}^2+|\Delta_{\mathbf k}|^2}$ where $\xi_{\mathbf k}=v_f(|\vk|-k_f)$ is the normal-state excitation energy measured from the Fermi level.
The Berry curvature reflects the topology of the order parameter, and thus decays rapidly away from the Fermi surface (c.f. Ref.~\cite{Schnyder:2015}). In the limit $|\Delta|\ll E_f$ the Berry curvature confines the summation over $\vk$ to the Fermi surface,
\begin{equation}
\lim_{|\Delta|/T\to0}\,\Omega_{k_x,k_y}^{(\pm 1)}(\mathbf{k})
=
\pm\nu k_f^{-1}\delta(|\mathbf{k}|-k_f)
\,,
\end{equation}
where as before $\nu$ is the phase winding of the order parameter about the $k_z$-axis. The resulting Berry phase contribution to the anomalous thermal Hall conductance for isotropic Fermi surfaces in $d$ dimensions is
\begin{equation}\label{eq:kappa_surface}
\kappa_{xy}^\text{edge}
=
-\frac{\nu}{2\pi\hbar T}
\left(\frac{ k_f}{\pi}\right)^{d-2}
\la\int_{-\Delta_{\p}}^{+\Delta_{\p}} d\e\,\e^2 f'(\e)\ra_\p 
=
\nu
\left(\frac{\pi k_B^2}{6\hbar}\right)
\left(\frac{k_f}{\pi}\right)^{d-2}
\,T 
\,,
\end{equation}
in the low-temperature limit. 

In Figs.~\ref{fig:K_zeroT_pointnode} and \ref{fig:K_zeroT_linenode} we compare the our results for the impurity-induced thermal Hall conductivity with the prediction of the edge contribution based on the Berry curvature in the low temperature limit ($T\to0$) for four different chiral ground states. The comparison is based on a typical coherence length scale $k_f\xi_0=100$.
For all four chiral states the Berry phase contribution is dominant at impurity densities below the threshold for impurity-induced transverse transport in the limit $T\to0$. However, above this threshold the impurity-induced Hall conductivity is comparable to or much larger than the Berry phase contribution. For example, for the chiral $E_{1u}$ state (bottom left panel of Fig.~\ref{fig:K_zeroT_pointnode}), the impurity-induced Hall effect yields transverse heat currents in the zero temperature limit which are approximately an order of magnitude larger than the Berry phase contribution for typical impurity dimensions.

\begin{figure}
\centering
\includegraphics[width=0.75\linewidth]{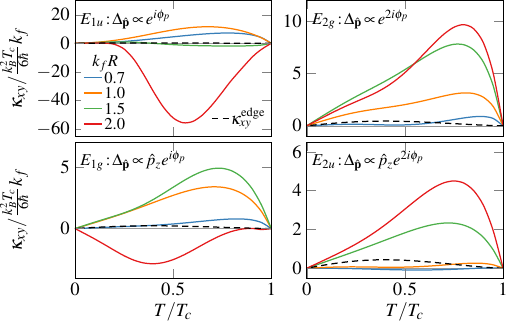}
\caption{%
The anomalous thermal Hall conductivity versus temperature for hard-sphere radii $k_fR\!=\!0.7,1,1.5,2.0$, and for chiral order parameters belonging to the 
$E_{1u}$, $E_{2g}$, $E_{1g}$ and $E_{2u}$
irreducible representations of the hexagonal point group.
The Berry phase contribution is shown for comparison (dashed curves).
Results are shown for a transport mean free path is $L_N/\xi_0\!=\!7.5$ and coherence length of $\xi_0\!=\!100\,k_f^{-1}$.
Figure reproduced from Ref.~\cite{wave:20} with permission of the APS and the authors.
}
\label{fig:Kxy_3d}
\end{figure}
%
Figure~\ref{fig:Kxy_3d} depicts the zero-field thermal Hall conductivity as a function of temperature for chiral states belonging to the spin-triplet, odd-parity $E_{1u}$ and $E_{2u}$ representations and the spin-singlet, even-parity $E_{1g}$ and $E_{2g}$ representations of the hexagonal $D_{6h}$ point group, and the $E_{u}$ and $E_{g}$ representations of $D_{4h}$. Almost all proposed chiral superconductor candidates, including the perovskite \sro\ and the heavy-fermion superconductor \upt, belong to one of these representations.
The results show that the impurity-induced anomalous Hall effect (solid lines) dominates the Berry curvature contribution~\cite{qin11,gos15} (dashed lines) over the full temperature range in all four chiral pairing states for impurities with $k_f R\gtrsim 1.5$. 
In this context it is worth reiterating our earlier estimate of the magnitude of the impurity-induced anomalous thermal Hall conductivity for the chiral phase of \upt~\cite{wave:20}. Namely for $k_f\!=\!1\,\text{\AA}^{-1}$, $\xi_0\!=\!100\,\text{\AA}$ and $T_c\!=\!0.5\,\text{K}$, representative of \upt~\cite{joy02} we estimate $\kappa_{xy}\!>\!3\!\times\!10^{-3}\,\text{WK$^{-1}$m$^{-1}$}$ for $T\simeq 0.75 T_c$ for the chiral $E_{2u}$ state with $\nu=2$ and impurity radius $k_fR\!=\!1.5$ (Fig.~\ref{fig:Kxy_3d}). Compared to the normal-state thermal conductivity at $T_c$, $\kappa_N(T_c)$, one needs sensitivity to transverse heat currents at the level of $0.01-0.03\,\kappa_N(T_c)$ as shown in Fig.~\ref{fig:kappa_xy_Tc}.\vspace*{-10mm}

\section{Summary and Outlook}

We presented the theoretical framework for understanding disorder-induced anomalous Hall transport in chiral superconductors, and we reported quantitative predictions for the thermal conductivity and the anomalous thermal Hall conductivity in superconductors with phase winding $\nu$ for chiral superconducting ground states belonging to the 2D irreducible representations of the hexagonal and tetragonal point groups. We highlight the role of quasiparticle-impurity scattering by \emph{finite-size impurities}, i.e. $k_f R \gtrsim 1$.
Our analysis demonstrates that an anomalous thermal Hall effect is obtained for chiral superconductors with winding $\nu$, provided the ionic radius of the impurities satisfies $k_f R \gtrsim |\nu|-1$.
Thus, for point-like impurities with $k_f R\ll 1$ the anomalous thermal Hall current vanishes for all but chiral p-wave ground states.
We also discussed the spectrum of impurity-induced Andreev bound states, which are formed via multiple Andreev scattering. The spectrum depends sensitively on the winding number of the chiral order parameter as well as the structure of the impurity potential. 
Our results also show that the impurity-induced anomalous thermal Hall transport dominates the edge state contribution by an order of magnitude or more over most of the temperature range below $T_c$.
The impurity- and edge contributions to the thermal Hall effect both depend on broken time-reversal and mirror symmetries. Thus, they are equally good signatures of chiral superconductivity. 
The bulk impurity effect is likely more accessible experimentally; it produces larger Hall currents, and it is insensitive to the quality of the surfaces of a sample.
In summary this work provides the theoretical framework for computing and analyzing experiments seeking to identify broken time-reversal and mirror symmetries, as well as non-trivial topology of chiral superconductors, from bulk transport measurements.

\noindent\emph{Outlook:} There are a number of candidates for chiral superconductivity that have been proposed theoretically and pursued experimentally. The chiral phase of \He\ was proven to be chiral p-wave based on the observation of anomalous Hall transport of electrons embedded in superfluid \Hea~\cite{ike13,she16}. 
The heavy electron metal \upt\ shows evidence of broken time-reversal symmetry based on Kerr rotation~\cite{sch14}, Josephson interferometry~\cite{str09}, $\mu$SR~\cite{luk93} and SANS studies of diffraction by the vortex lattice~\cite{avers:20}. Observation of an anomalous thermal Hall effect onsetting at the A to B transition would provide a definitive bulk signature of broken time-reversal and mirror symmetries in \upt. Analysis of the temperature- and impurity-dependences of the Hall conductivity could provide new and quantitative experimental constraints on the symmetry class of E-rep of \upt. 
For a number of proposed candidates for chiral superconductivity, e.g.\ \sro, doped graphene, SrPtAs, etc., observation of an anomalous thermal Hall effect would provide confirmation of broken time-reversal and mirror symmetry by the superconducting order parameter.
NMR experiments revealed the existence of new superfluid phases of liquid \He\ when it is infused into low density, anisotropic, random solids - ``aerogels''~\cite{pol12} - or confined into sub-micron cavities~\cite{lev13}. Analysis based on Ginzburg-Landau theory predicts that the ground state of \He\ under anisotropic confinement is a chiral phase~\cite{sau13}. Thus, experiments designed to measure the transverse heat current could provide a definitive test of the theory for the ground state of superfluid \He\ infused into anisotropic aerogels~\cite{sha22}, and similarly for \He\ confined in sub-micron cavities~\cite{sha22a}.

\paragraph*{Conflict of Interest Statement}
The authors declare that the research was conducted in the absence of any commercial or financial relationships that could be construed as a potential conflict of interest.
\paragraph*{Author Contributions}
Both authors contributed to all aspects of this work.
%
\paragraph*{Funding}
The research of VN was supported through the Center for Applied Physics and Superconducting Technologies at Northwestern University and Fermi National Accelerator Laboratory. The research of JAS was supported in part by the National Science Foundation (Grant DMR-1508730),``Nonequilibrium States of Topological Quantum Fluids and Unconventional Superconductors'', and by the U.S. Department of Energy, Office of Science, National Quantum Information Science Research Centers, Superconducting Quantum Materials and Systems Center (SQMS) under contract number DE-AC02-07CH11359.

\paragraph*{Acknowledgments}
We thank Pallab Goswami for discussions that informed this work.
\vfill\eject
%
%
\end{document}